\documentclass[12pt]{article}
\usepackage[top=2.828cm,bottom=2.828cm,left=2.1cm,right=2.1cm]{geometry}
\usepackage{amsmath,amssymb}
\usepackage{amsthm}
\usepackage{mathrsfs}
\usepackage{xcolor}
\numberwithin{equation}{section}

\newcommand{\na}{\nabla}
\renewcommand{\b}{\bar}
\renewcommand{\d}{\dot}

\newcommand{\co}{{\rm c}}
\newcommand{\bs}{\boldsymbol}
\newcommand{\df}{\dfrac}
\newcommand{\ul}{\underline}
\newcommand{\der}{\partial}
\renewcommand{\(}{\left(}
\renewcommand{\)}{\right)}
\newcommand{\scr}{\mathscr}
\renewcommand{\cal}{\mathcal}
\newcommand{\bmx}{\left(\begin{matrix}}
\newcommand{\emx}{\end{matrix}\right)}
\newcommand{\corr}{\leftrightarrow}
\newcommand{\axg}{
\bmx
1 & 0 \\ 0 & -1
\emx
} 
\newcommand{\stog}[4]{
\bmx
0 & \!(\sigma_{#1}^{#2})_{{#3}\dot{#4}} \\
(\bar{\sigma}_{#1}^{#2})^{\dot{#3}{#4}} \! & 0
\emx 
}
\newcommand{\stogm}[4]{
\bmx
0 & \!(\sigma_{#1}^{#2})_{{#3}\dot{#4}} \\
-(\bar{\sigma}_{#1}^{#2})^{\dot{#3}{#4}} \! & 0
\emx 
}
\newdimen\Tdim
\newdimen\Ddim
\def\Tspan#1{{\setbox0=\hbox{$#1$}%
\Tdim\ht0\advance\Tdim\dp0\advance\Tdim.7ex\Ddim\dp0\advance\Ddim.4ex\rule[-\Ddim]{0pt}{\Tdim}\box0}}

\newcommand{\Sqrbra}[4]{\raise-#4\hbox{
\rule{#3}{#1}\hskip-#3
\rule{#2}{#3}\hskip-#2
\rule[#1]{#2}{#3}}
\hskip-#2
}
\newcommand{\Sqrket}[5]{\raise-#4\hbox{
\rule{#2}{#3}\hskip-#2
\rule[#1]{#2}{#3}
\rule[#1]{#3}{#3}
\hskip-#5
\rule{#3}{#1}}}

\def\cfbra{\Sqrbra{14pt}{4.1pt}{0.45pt}{3.9pt}
\kern-.61pt\big(}
\def\Bigbra{\Sqrbra{21pt}{5.5pt}{0.5pt}{7.8pt}
\kern-.7pt\Big(}
\def\biggbra{\Sqrbra{28.3pt}{7pt}{0.55pt}{11.5pt}
\kern-0.85pt\bigg(}
\def\cfket{
\big) \kern-5.45pt\Sqrket{14pt}{4.05pt}{0.45pt}{3.9pt}{4.54pt}}
\def\Bigket{
\Big) \kern-6.5pt\Sqrket{21pt}{5.45pt}{0.5pt}{7.8pt}{4.8pt}}
\def\biggket{
\bigg) \kern-7.8pt\Sqrket{28.3pt}{6pt}{0.55pt}{11.5pt}{4.5pt}}

\def\nn{\nonumber\\}

\begin{document}
\begin{titlepage}

\begin{flushright}
\hfill MISC-2016-01\\
\end{flushright}

\vspace{10mm}

\begin{center}
{\Large\bf%
Component versus Superspace Approaches\\[3mm]%
to $D=4$, $\cal{N}=1$ Conformal Supergravity}
\vspace{12mm}

{\large%
Taichiro Kugo$^1$, Ryo Yokokura$^2$ and Koichi Yoshioka$^3$}\\
\vspace{5mm}

$^1${\it Department of Physics and Maskawa Institute for Science 
and Culture,}\\
{\it Kyoto Sangyo University, Kyoto 603-8555, Japan}\\
$^2${\it Department of Physics, Keio  University, Yokohama 223-8522, Japan}\\
$^3${\it Osaka University of Pharmaceutical Sciences, Takatsuki 
569-1094, Japan}

\vspace{15mm}
{\bf Abstract}

\end{center}

\noindent
The superspace formulation of $\cal{N}=1$ conformal supergravity in
four dimensions is demonstrated to be equivalent to the conventional
component field approach based on the superconformal tensor
calculus. The detailed correspondence between two approaches is
explicitly given for various quantities; superconformal gauge fields,
curvatures and curvature constraints, general conformal multiplets and
their transformation laws, and so on. In particular, we carefully
analyze the curvature constraints leading to the superconformal
algebra and also the superconformal gauge fixing leading to Poincar\'e
supergravity since they look rather different between two approaches.

\end{titlepage}
\setcounter{footnote}{0}

\section{Introduction}

$\cal{N}=1$ supergravity (SUGRA) in four dimensions has been important
as giving a boundary theory around the unification scale for
constructing viable phenomenological models beyond the standard
model. It has also become to have increasing importance as low-energy
effective theory of superstring and as a tool for analyzing
supersymmetric gauge theories on curved backgrounds.

However various explicit calculations, e.g., the construction of SUGRA
Lagrangian, are complicated and non-trivial. The simplest and most
convenient method is presumably the superconformal tensor calculus,
which was developed by Kaku, Townsend, van Nieuwenhuizen, Ferrara,
Grisaru, de Wit, van Holten and 
Van Proeyen~\cite{bib:KTVN}-\cite{bib:dWvHVP2}. It is a set of rules
for constructing invariant actions under local superconformal
transformations; that is, superconformal gauge fields including
gravity and gravitino and various types of matter multiplets, their
transformation laws, multiplication rules, and superconformal
invariant action formulas. The power of the superconformal tensor
calculus comes from larger symmetry than the usual Poincar\'e
SUGRA\@. Indeed its power as a practical computational tool was clearly
demonstrated in Ref.~\cite{bib:KUigc} for computing the action for
the general Yang-Mills-matter coupled SUGRA system.

Kugo and Uehara (KU) have presented~\cite{bib:KU} the superconformal
tensor calculus in the most complete form, and discussed the spinorial
derivative $\scr{D}_{\ul{\alpha}}$ for the first time in the component
field approach. They found that a special condition on an operand
multiplet $\cal{V}_\Gamma$ must be satisfied so that its spinorial 
derivative $\scr{D}_{\ul{\alpha}}\cal{V}_\Gamma$ exists and gives a
conformal multiplet. The condition depends on the spinor index
$\ul{\alpha}$ of $\scr{D}_{\ul{\alpha}}$ and the Lorentz index
$\Gamma$ of the operand $\cal{V}_\Gamma$, and KU implicitly suspected
that the superspace formulation might not exist for the conformal SUGRA.

Nevertheless Butter \cite{bib:B1} has recently presented a superspace
formalism of the conformal SUGRA\@. Contrary to the previous
expectation, his formalism realizes a simpler algebra of covariant
derivatives than any other superspace Poincar\'e SUGRA:
\begin{equation}
\{\na_\alpha, \na_\beta \} = \{\b\na_{\d\alpha},\b\na_{\d\beta}\}=0, \qquad 
\{\na_\alpha,\b\na_{\d\beta}\} = -2i\na_{\alpha\d\beta}.
\end{equation}
Requiring this algebra together with several constraints on curvatures
in the vector-spinor direction, he succeeded in constructing a
superspace counterpart of the conformal SUGRA in component
approach. The covariant derivatives 
$\na_A=(\na_a,\,\na_\alpha,\,\b\na^{\d\alpha})$ can be freely
applied on any superfield with no restriction and are identified with
the transformations $P_A=(P_a,\,Q_\alpha,\,\b{Q}^{\d\alpha})$ of
superconformal group. The reason why KU's spinorial derivatives could
not be freely applied turns out that KU required an extraneous
condition that the derivative again give a primary multiplet.

Since the superspace formalism manifests supersymmetry in a
geometrically clear way, it gives transparent and powerful means to
treat the systems in new situations such as finding non-linear
realization, brane world, decomposition of higher $\cal{N}$
supersymmetry, partial breaking of local supersymmetry, massive SUGRA,
etc. On the other hand, one needs to write down the action explicitly
in terms of component fields, which could be done most easily and
efficiently with the tensor calculus. That is, we have two approaches
to the conformal SUGRA, one is the superspace approach based on the
conformal superspace and superfields, and the other is the component
approach based on the superconformal tensor calculus. Both approaches
have their own strong and weak points. In order to use the advantages
of both approaches, it is desirable to see the correspondence between
them. The purpose of this paper is to show the equivalence of two
approaches by making the detailed correspondences manifest.

This paper is organized as follows. In section~2, we recapitulate the
essential parts, first, of the superconformal tensor calculus in
component approach, and then, of the conformal superspace approach. We
use the individual notation for each of these approaches and
separately give a dictionary between them for the convenience of
reading the references.

In section~\ref{sec:corr} we explicitly present the correspondences of 
various quantities. We first discuss gauge fields and curvatures in
Sec.~3.1 and show how all the curvature constraints in component
approach are satisfied in superspace approach, although the
constraints look rather different from each other. The same
superconformal transformation algebras are realized in both approaches
under these curvature constraints. We then discuss the component
fields and transformation rules for a conformal multiplet with
arbitrary external Lorentz index in Sec.~3.2, and the chiral
projection and the invariant action formulas in Sec.~3.3. We analyze
in Sec.~3.4 the compensated (or $u$-associated) derivatives which map
a primary superfield to primary one. There we also discuss the KU's
restriction on the spinorial derivatives.

In section~\ref{sec:scym}, we investigate the matter-coupled SUGRA
system and the superconformal gauge fixing to Poincar\'e SUGRA, mainly
from the superspace viewpoint. We discuss the superspace counterpart
of the KU's gauge fixing which leads directly to the canonically
normalized Einstein-Hilbert (EH) and Rarita-Schwinger (RS) terms. The
correspondence to the component approach is non-trivial since the
gauge invariance in superspace approach is much larger than the
component approach, and the gauge fixing written in terms of
superfields give more fixing conditions than the component case. One
remarkable fact is that the covariant spinor derivatives remaining
after the gauge fixing automatically reproduce the complicated
supersymmetry transformation in Poincar\'e SUGRA\@. The final section
is devoted to the summary. We add three appendices. The notations in
the component and superspace approaches are summarized separately and
the dictionary between them is given in appendix A\@. The standard
form of supersymmetry transformation law for the general conformal
multiplet with arbitrary external Lorentz index is cited for
convenience in appendix B\@. We present in appendix C some explicit
computations which are necessary in deriving the results in the text.

\section{Conformal SUGRA}

We first briefly review the component and superspace approaches for
$D=4$, $\cal{N}=1$ conformal SUGRA\@. In both approaches the conformal
SUGRA is constructed as the gauge theory of superconformal group. The
Lie algebra of the superconformal group contains the following
elements: translation $P_a$, supersymmetry $Q$, Lorentz
transformation $M_{ab}$, conformal boost $K_a$, supersymmetry of
conformal boost $S$, dilatation $D$ and chiral rotation $A$.

\subsection{Component approach}
\label{sec:comapp}

In this subsection we review the component approach.
For the component approach part in this paper, 
we use the notations and conventions of Ref.~\cite{bib:KU}, 
which are the same as those of Ref.~\cite{bib:vN} except for
two-component spinors and the dual of anti-symmetric tensors. The
detail of the notations is summarized in
appendix~\ref{sec:notation}. The superconformal algebra consists of 
15 bosonic and 8 fermionic generators, which
obey the following graded commutation relations:
\begin{equation}
\begin{split}
& [M_{ab},M_{cd}]=-M_{ad}\delta_{bc}+M_{bd}\delta_{ac}+M_{ac}\delta_{bd}
-M_{bc}\delta_{ad},\\
& [M_{ab},P_c]=-P_a\delta_{bc}+P_b\delta_{ac},\qquad 
[M_{ab},K_c]=-K_a\delta_{bc}+K_b\delta_{ac}, \\
& [D,P_a]=P_a,\qquad [D,K_a]=-K_a,
\qquad [K_a,P_b]=2\delta_{ab}D+2M_{ab}, \\ 
& \{Q,Q^T\}=-\df{1}{2}(\gamma_aC^{-1})P_a,
\qquad \{S,S^T\}=\df{1}{2}(\gamma_a C^{-1})K_a, \\
& [M_{ab},Q]=\sigma_{ab}Q,\qquad [M_{ab},S]=\sigma_{ab}S, \\
& [D,Q]=\df{1}{2}Q,\quad [D,S]=-\df{1}{2}S,
\qquad [A,Q]=-\df{3}{4}i\gamma_5 Q,\quad [A,S]=\df{3}{4}i\gamma_5 S, \\
&[K_a,Q]=-\gamma_a S,\qquad [S,P_a]=-\gamma_a Q, \\
& \{Q,S^T\} =
-\df{1}{2}C^{-1}D+\df{1}{2}\sigma^{ab}C^{-1}M_{ab}+i\gamma_5C^{-1}A.
\end{split} 
\label{eq:SCalgebra}
\end{equation} 
All other commutation relations vanish. The generators are generically
denoted as $X_A$ and the above commutation relations are written as
\begin{equation}
[X_A,X_B\}=-{f_{AB}}^C X_C. 
\end{equation}
Note that these generators represent the active operators transforming
fields, not the representation matrices. The commutation relations
change the signs if written for representation matrices instead of
active operators. In the conformal SUGRA, the superconformal symmetry
is treated as local symmetry. The corresponding gauge fields and
transformation parameters are given by
\begin{align}
 {h_\mu}^A X_A &= {e_\mu}^a P_a
+\b\psi_\mu Q +\df{1}{2}{\omega_\mu}^{ab}M_{ab}
+b_\mu D+A_\mu A+\b\varphi_\mu S +{f_\mu}^a K_a, \\
 \epsilon^A X_A &= \xi^a P_a
+\b\varepsilon Q +\df{1}{2}\lambda^{ab}M_{ab} 
+\rho D+\theta A +\b\zeta S +{\xi_K}^a K_a.
\end{align}
In component approach, the Greek letters $\mu,\nu,...$ denote the
curved vector indices and the Roman letters $a,b,...$ the flat Lorentz
indices. The group transformation laws of the gauge fields under the
superconformal symmetry are
\begin{equation}
\delta^{\text{group}}_B(\epsilon^B){h_\mu}^A = \der_\mu \epsilon^A 
+ {h_\mu}^B\epsilon^C {f_{CB}}^A.
\end{equation}
The curvature of the superconformal algebra (before the deformation
below) is
\begin{equation}
R_{\mu\nu}^A = \der_\nu {h_\mu}^A -\der_\mu{h_\nu}^A
+{h_\nu}^B {h_\mu}^C {f_{CB}}^A.
\end{equation}
The $P_a$ translation is deformed so as to be related to the
general coordinate (GC) transformation $\delta_{\text{GC}}$ as 
\begin{equation}
\delta_{\tilde{P}}(\xi^a)
:=\delta_{\text{GC}}(\xi^\mu)-\sum_{A\neq P} \delta_A(\xi^\mu{h_\mu}^A),
\end{equation}
where $\xi^\mu=\xi^ae_a{}^\mu$, and $\xi^a$ is a field-independent
parameter. In order to have $[\delta_Q,\delta_Q] \sim \delta_{\tilde{P}}$, 
several constraints on the curvatures are imposed:
\begin{equation}
{R_{\mu\nu}(P^a)} = 0,
\label{eq:constP}
\end{equation}
\begin{equation}
R_{\mu\nu}(Q)\gamma^\nu = 0,
\label{eq:constQ}
\end{equation}
\begin{equation}
R_{\nu\lambda}(M_{ab})e^{a\lambda}{e^{b}}_{\mu}
-\df{1}{2}R_{\lambda\mu}(Q)\gamma_\nu \psi^\lambda 
+\df{1}{2}i\tilde{R}_{\mu\nu}(A) = 0,
\label{eq:constM}
\end{equation}
where $\tilde{R}_{\mu\nu}$ is the dual of $R_{\mu\nu}$. By these
constraints, the $M_{ab}$, $S$ and $K_a$ gauge fields 
($\omega_\mu{}^{ab}$, $\varphi_\mu$ and $f_\mu{}^a$,  respectively)
become dependent fields expressed by other independent gauge
fields. The $Q$ transformations $\delta_Q(\varepsilon)$ of the
dependent gauge fields are determined by those of independent gauge
fields, and they deviate from the original group transformation 
$\delta_Q^{\rm group}(\varepsilon)$ as
\begin{equation}
\delta_Q(\varepsilon) = \delta^{\rm group}_Q(\varepsilon)
+\delta'_Q(\varepsilon).
\end{equation}
The deviation part $\delta'_Q(\varepsilon)$ is given by
\begin{equation}
\begin{split}
\delta'_Q(\varepsilon){\omega_\mu}^{ab} &=
\df{1}{2}R^{ab}(Q)\gamma_\mu \varepsilon, \\
\delta'_Q(\varepsilon)\varphi_\mu &=
\df{1}{4}i\gamma^\nu(\gamma_5 R_{\nu\mu}(A)+
\tilde{R}_{\nu\mu}(A))\varepsilon, \\
\delta'_Q(\varepsilon){f_\mu}^a &=
-\df{1}{2}R^{\text{cov}}_{\nu\mu}(S)\sigma^{a\nu}\varepsilon
-\df{1}{4}e^{a\nu}\tilde{R}^{\text{cov}}_{\nu\mu}(S)\gamma_5\varepsilon,
\end{split}
\label{eq:Qprime}
\end{equation}
where
\begin{equation}
{R^{\text{cov}}_{\mu\nu}}^A = {R_{\mu\nu}}^A+\delta'_Q(\psi_\mu)h_\nu^A
-\delta'_Q(\psi_\nu){h_\mu}^A.
\end{equation}
Note that the RHS of Eq.~(\ref{eq:Qprime}) are given by 
$\varepsilon e_\mu^c f_{Q\,P_c}{}^X$ with $X=M_{ab},\, S,\, K_a$. So
they can be regarded as the deformation of the algebra by changing the
structure constant of $[Q,\,P_c]$ commutator from (originally) zero to
the non-vanishing $f_{Q\,P_c}{}^X$ for $X=M_{ab},\, S,\, K_a$.

The resultant commutation relations are the same as the original ones 
\begin{equation}
[\delta_A(\epsilon_1^A),\delta_B(\epsilon_2^B)]
=\sum_C \delta_C(\epsilon_1^A\epsilon_2^B{f_{BA}}^C),
\end{equation}
for all $A$ and $B$, if $P_a$ on the RHS of $Q$-$Q$ commutator is
understood to be $\tilde P_a$:
\begin{equation}
[\delta_Q(\varepsilon_1),\delta_Q(\varepsilon_2)] = \delta_{\tilde{P}}
\Bigl(\df{1}{2}\b\varepsilon_2\gamma^a\varepsilon_1\Bigr).
\end{equation}
Moreover, the definition of $\tilde P_a$ transformation leads to
\begin{align}
[\delta_{\tilde{P}}(\xi^a),\delta_Q(\varepsilon)] &= 
\sum_{A=M,S,K} \delta_A(\xi^a\delta'_Q(\varepsilon){h_a}^A) =
\sum_{A=M,S,K} \delta_A(\xi^a\varepsilon^{\underline\alpha} 
f_{Q_{\underline\alpha}P_a}{}^A ),  \nonumber \\
[\delta_{\tilde{P}}(\xi_1^a),\delta_{\tilde{P}}(\xi_2^b)] &=
\;\sum_{A\neq P} \delta_A(\xi_1^a\xi_2^b{R^{\text{cov}}_{ab}}^A), 
\qquad \(\to\, f_{P_aP_b}{}^A = -{R^{\text{cov}}_{ab}}^A \)
\end{align}
where $\ul{\alpha}=(\alpha,\d\alpha)$. The superconformally covariant
derivative on fields carrying only flat Lorentz indices is defined
through the $\tilde{P}_a$-transformation as
\begin{equation}
\xi^a D_a\phi := \delta_{\tilde{P}}(\xi^a)\phi
=\xi^a{e_a}^\mu\der_\mu \phi-\sum_{A\neq \tilde{P}}\delta_A(h_a{}^A)\phi. 
\label{eq:cfcovder}
\end{equation}

Next, we introduce superconformal multiplets. A general conformal
multiplet $\cal{V}_\Gamma$ is a set of $(8+8)\times \dim\Gamma$
complex fields,
\begin{equation}
\cal{V}_\Gamma = [\cal{C}_\Gamma,\,\cal{Z}_{\ul{\alpha}\Gamma},\,
\cal{H}_\Gamma,\, \cal{K}_\Gamma,\, \cal{B}_{a\Gamma},\, 
\Lambda_{\ul{\alpha}\Gamma},\, \cal{D}_\Gamma], 
\label{eq:cfmatter}
\end{equation}
where $\Gamma$ represents arbitrary spinor indices 
$\Gamma=(\alpha_1,...,\alpha_m; \d\beta_1,...,\d\beta_n)$ and 
$\dim\Gamma$ is the dimension of Lorentz representation of
$\Gamma$. The first component $\cal{C}_\Gamma$ is defined to have the
lowest Weyl weight in the multiplet so that its transformation law is
given by
\begin{equation}
\begin{split}
& \delta_Q(\varepsilon)\cal{C}_\Gamma = \df{1}{2}i\b{\varepsilon}\gamma_5
\cal{Z}_\Gamma, \qquad 
\delta_M(\lambda^{ab})\cal{C}_\Gamma =
\df{1}{2}\lambda^{ab}{(\Sigma_{ab})_\Gamma}^\Sigma \cal{C}_\Sigma
=: \df{1}{2}\lambda^{ab}(\Sigma_{ab}\cal{C})_\Gamma  \\ 
& \bigl(\delta_D(\rho) + \delta_A(\theta)\bigr) \cal{C}_\Gamma = 
\Bigl(w\rho + \df{1}{2}in\theta \Bigr)\cal{C}_\Gamma, \qquad
\delta_S(\zeta)\cal{C}_\Gamma = \delta_K(\xi_K^a) \cal{C}_\Gamma=0.
\end{split} 
\label{eq:tlol}
\end{equation}
Here $\Sigma^{ab}$ is the representation matrix of Lorentz generator
which $\cal{C}_\Gamma$ belongs to, and $w$ and $n$ are the Weyl and
chiral weights of $\cal{C}_\Gamma$. The $S$ and $K_a$ transformations
must annihilate the lowest weight component $\cal{C}_\Gamma$ since
they lower the Weyl weights of operands. The $Q$ transformation law
$\delta_Q(\varepsilon)\cal{C}_\Gamma=
\frac{1}{2}i\b{\varepsilon}\gamma_5\cal{Z}_\Gamma$ simply defines the
second component $\cal{Z}_\Gamma$. All the higher components in the
multiplet and their superconformal transformation laws are determined
by demanding the superconformal algebra to hold on them, aside from
some arbitrariness in defining higher component fields. The $Q$
transformation laws of all component fields are summarized in
(\ref{eq:sf}), which also fix the definition of higher component 
fields. We call the transformation laws (\ref{eq:sf}) the standard
form. Since the first component ${\cal C}_\Gamma$ specifies the whole
multiplet, we denote the conformal multiplet $\cal{V}_\Gamma$ using
the first component as
\begin{equation}
\cal{V}_\Gamma = \cfbra \cal{C}_\Gamma \cfket \,.
\end{equation}

A constrained-type multiplet also exists as a 
conformal multiplet if some conditions are met on Weyl and chiral 
weights and also on its Lorentz representation. The chiral multiplet
$\Sigma^{(w,n)}_\Gamma$, for instance, exists only when the Weyl and
chiral weights $(w,n)$ satisfy $w=n$ and the Lorentz index $\Gamma$ is 
made of purely undotted spinor indices; then the chiral multiplet 
has $(2+2)\times\dim\Gamma$ complex components denoted by
\begin{equation}
\Sigma^{(w=n)}_{\Gamma=(\alpha_1\cdots\alpha_l)}=\left[
\cal{A}_\Gamma,\ 
\cal{P}_\text{R}\chi_\Gamma ,\
\cal{F}_\Gamma\right].
\label{eq:chiral_comp}
\end{equation} 
These three components of a chiral multiplet are embedded into a
general conformal multiplet in the form $\cal{V}(\Sigma_\Gamma)
=[\cal{A}_\Gamma,-i\cal{P}_\text{R}\chi_\Gamma,-\cal{F}_\Gamma,i
\cal{F}_\Gamma,iD_a\cal{A}_\Gamma,0,0]$, so that their $Q$ and $S$
transformation laws are given by
\begin{align}
\delta_{QS}\cal{A}_\Gamma &=
\bigl(\delta_Q(\varepsilon)+\delta_S(\zeta)\bigr)\cal{A}_\Gamma 
=\frac12 \b\varepsilon_\text{R} \cal{P}_\text{R} \chi_\Gamma,  \nn
\delta_{QS} \cal{P}_\text{R} \chi_\Gamma &= 
(-1)^\Gamma\Big( \gamma^a D_a\cal{A}_\Gamma\varepsilon_{\rm L}
+\cal{F}_\Gamma\varepsilon_\text{R} +\(2w\cal{A}_\Gamma 
-(\Sigma^{ab}\cal{A})_\Gamma\sigma_{ab}\)\zeta_\text{R} \Big),
\label{eq:222}  \\
\delta_{QS}\cal{F}_\Gamma &=
\frac12 \b\varepsilon_{\rm L}\gamma^a D_a \cal{P}_\text{R} \chi_\Gamma
+\bar\zeta_\text{R}\Bigl( (1-w)\cal{P}_\text{R} \chi_\Gamma
-\frac12\sigma_{ab} (\Sigma^{ab} \cal{P}_\text{R} \chi )_\Gamma 
\Bigr). \nonumber
\end{align}
For the multiplet $\cal{V}^{(w,n)}_\Gamma$ with purely undotted 
spinor $\Gamma$ satisfying $w=n+2$, the chiral projection operator
$\Pi$ exists and
\begin{equation}
\Pi \cal{V}^{(w,\,n=w-2)}_\Gamma = 
\Big[ \, \df{1}{2}(\cal{H}_\Gamma-i\cal{K}_\Gamma),\ 
i\cal{P}_{\text{R}}(\gamma^a D_a\cal{Z}_\Gamma+\Lambda_\Gamma),\ 
-\df{1}{2}(\cal{D}_\Gamma +
\square \cal{C}_\Gamma+iD^a \cal{B}_{a\Gamma})\,\Big]
\label{eq:chiralproj}
\end{equation}
gives a chiral multiplet with the Weyl and chiral weights
$(w+1,w+1)$. Here $\square=D^aD_a$ is the superconformal d'Alembertian.

The superconformal tensor calculus gives the superconformally
invariant action in simple forms. The F-type invariant action formula
is applied only to the chiral multiplet $\Sigma=\bigl[\,\cal{A}=
\frac{1}{2}(A+iB),\, \cal{P}_\text{R}\chi,\, \cal{F}=
\frac{1}{2}(F+iG)\,\bigr]$ satisfying $w=n=3$ and carrying no external
Lorentz index. The action is given by 
\begin{equation}
\int d^4 x \bigl[\,\Sigma^{(w=n=3)}\,\bigr]_F
=\int d^4 x \, e \Bigl(F+\df{1}{2}\b\psi_a \gamma^a \chi + 
\df{1}{2}\b\psi_a\sigma^{ab}(A-i\gamma_5 B)\psi_b\Bigr).
\label{eq:F-action}
\end{equation}
The D-type invariant action formula is applied only to the real and
Lorentz-scalar multiplet $V=\bigl[\,C,\,Z,\,H,\,K,\,B_a,\,\lambda,\,
D\,\bigr]$ with $w=2$ and $n=0$. The action is derived from the F-type
formula with the chiral projection operator $\Pi$ as 
\begin{align}
& \int d^4x \bigl[\,V^{(w=2)}\,\bigr]_D = 
\int d^4 x \bigl[\,-\Pi V^{(w=2)}\,\bigr]_F  \nn
& = \int d^4 x \,e\Bigl(\,D+\square C -\df{1}{2}i\b\psi_a\gamma^a\gamma_5
(\gamma^b D_b Z+\lambda) -\df{1}{2}\b\psi_a
\sigma^{ab}(H+i\gamma_5K)\psi_b \,\Bigr)  \nn
& = \int d^4 x\, e \( \,D-\df{1}{2}i\b\psi_a\gamma^a\gamma_5 \lambda
-i\b\varphi_a \gamma^a\gamma_5 Z + 
\df{1}{3}C\Bigl(R+\df{1}{e}\b\psi_\mu 
\varepsilon^{\mu\nu\rho\sigma}\gamma_5\gamma_\nu 
\Big(\der_\rho\psi_\sigma +
\frac{i}{4}{\omega_\rho}^{ab}\sigma_{ab}\psi_\sigma\Big)\Bigr)\right. \nn
& \left.  \hspace{10em}{}+\df{1}{4}i\varepsilon^{abcd}\b\psi_a
\gamma_b \psi_c \Bigl(B_d-A_d C -\df{1}{2}\b\psi_d Z\Bigr) \,\).
\label{eq:D-action}
\end{align}

For the general YM-matter coupled SUGRA system, the action is given by
\begin{equation}
\begin{split}
\cal{L} &= -\df{1}{2}\bigl[\,\tilde\phi(S,\b{S} e^{2V_\text{G}})
\Sigma_{\text{c}}\b\Sigma_{\text{c}}\,\bigr]_D
+ \left[\,\Sigma_{\text{c}}^3g(S)\,\right]_F
-\df{1}{4}\bigl[\,f_{\alpha\beta}\b{W}^\alpha W^\beta\,\bigr]_F  \\
&= -\df{1}{2}\bigl[\,\phi(S,\b{S}e^{2V_\text{G}})\Sigma_0\b\Sigma_0\,\bigr]_D
+\bigl[\,\Sigma_0^3\,\bigr]_F
-\df{1}{4}\bigl[\,f_{\alpha\beta}\b{W}^\alpha W^\beta\,\bigr]_F ,
\end{split}
\end{equation}
where $S_i=[z_i,\cal{P}_{\text{R}}\chi_i,h_i]$ are the chiral matter
multiplets with vanishing weights $w=n=0$ and $\b{S}^i$ are their
conjugate. In the first term, $V_{\text{G}}$ means the YM vector
multiplet of internal symmetry. The field $\Sigma_\co$ is a chiral
compensator carrying weights $(w,n)=(1,1)$. For the system possessing
non-vanishing superpotential $g(S)$, it is convenient to redefine the
compensator as $\Sigma_\co\rightarrow\Sigma_0=g^{1/3}(S)\Sigma_\co
=[z_0,\cal{P}_\text{R}\chi_0,h_0]$ so that $\phi$ becomes the
combination of $\tilde\phi$ and superpotential: 
$\phi(S,\b{S}e^{2V_\text{G}})
=\tilde\phi(S,\b{S}e^{2V_\text{G}})|g(S)|^{-2/3}$. In the third term,
$f_{\alpha\beta}$ is a holomorphic functions of $S_i$, symmetric
under the exchange $\alpha \leftrightarrow \beta$, and $W^\alpha$ is
the gaugino multiplet (field-strength supermultiplet) of internal
symmetry. For the YM vector multiplet, the Wess-Zumino (WZ) gauge is
imposed, and then the gaugino multiplet is constructed by the $Q$
transformation that preserves the WZ gauge. We denote such $Q$
transformation as $\delta_Q^{\text{YM}}(\varepsilon)$.

To go down to the Poincar\'e SUGRA, we fix the extraneous 
$D$, $A$, $S$, $K_a$ gauge symmetries. 
The so-called {\it improved gauge-fixing conditions} adopted in
\cite{bib:KUigc} are  
\begin{equation}
\begin{split}
D,\ A\hbox{-gauge}&:\ z_0 = z^*_0 = \sqrt{3} \phi^{-\frac{1}{2}}(z,z^*), \\
S\hbox{-gauge}&:\ \chi_{\text{R}0}=-z_0\phi^{-1}\phi^i\chi_{\text{R}i},
\qquad K_a\hbox{-gauge}:\ b_\mu=0,
\label{eq:KUgaugeCond}
\end{split}
\end{equation}
where $\chi_{\text{R}0}=\frac{1}{2}\cal{P}_\text{R}\chi_0$ and
$\chi_{\text{R}i}=\frac{1}{2}\cal{P}_\text{R}\chi_i$. These gauge
conditions set the first and second components of the vector multiplet
$\phi\Sigma_0\b\Sigma_0$ to 3 and 0, respectively, in the D-type
action formula. As a result, the canonically normalized EH and RS
terms are obtained directly.

The relation between the $Q$ transformation 
$\delta_Q^\text{P}(\varepsilon)$ in the resultant Poincar\'e SUGRA and
the gauge-fixed conformal $Q$ transformation is given by
\begin{equation}
\delta_Q^\text{P}(\varepsilon) =
\delta^\text{YM}_Q(\varepsilon) +\delta_A(\theta(\varepsilon)) +
\delta_S(\zeta(\varepsilon)) +\delta_K(\xi^a(\varepsilon)),
\label{eq:omPQ}
\end{equation}
where
\begin{equation}
\begin{split}
\theta(\varepsilon) &=
-\df{i}{3}\(\cal{G}^i\b\varepsilon_\text{R}\chi_{\text{R}i}
-\cal{G}_i \varepsilon_\text{L}\chi_{\text{L}}^i \),  \\
\zeta_\text{R}(\varepsilon) &=
-\df{1}{2}\Bigl(h_0z_0^{-1}+\frac{1}{3}h_i\cal{G}^i\Bigr)\varepsilon_\text{R} 
-\df{1}{3}\(\Bigl(\cal{G}^{ij}-\df{1}{3}\cal{G}^i\cal{G}^j\Bigr)
\b\varepsilon_\text{R}\chi_{\text{R}j}
+\cal{G}^i_j \b\varepsilon_\text{L}\chi_{\text{L}}^j \)\chi_{\text{R}i}  \\
& \quad\qquad 
-\df{1}{12}\(\cal{G}^i\gamma^a\na_a z_i-\cal{G}_i\gamma^a\na_a z^{*i}\)
\varepsilon_\text{L} +\df{1}{4}i\gamma^a A_a\varepsilon_\text{L},  \\
\xi_a(\varepsilon) &=
\df{1}{4}\( \b\varphi_a\varepsilon -\b\psi_a\zeta(\varepsilon) \).
\end{split}
\label{eq:PQparameters}
\end{equation}
In this expression $\na_\mu z_i$ is the covariant derivative of
the internal symmetry, and $\cal{G}$ are given by
$\cal{G}=3\log\frac{1}{3}\phi(z,z^*)$. The indices of $\cal{G}$
represent the differentiation with respect to $z_i$ and $z^{*\,i}$, 
e.g., $\cal{G}^i{}_j=\partial^2\cal{G}/ \partial z_i\partial z^{*\,j}$.

\subsection{Conformal superspace}
\label{sec:conf_ss}

Next we review the conformal superspace approach~\cite{bib:B1}. In
superspace, the supersymmetry transformation can be treated as a
translation in the direction of the Grassmannian spinor coordinate on
the same footing as the usual translation $P_a$. The
(anti-)commutation relations between the spinor covariant derivatives
become complicated in Poincar\'e SUGRA, whereas in conformal
superspace, they are as simple as in global supersymmetry. In the
superspace approach part in this paper, we use notations and
conventions of Butter~\cite{bib:B1} with a few exceptions which will
be explained below. The detail of the notations is summarized
in appendix~\ref{sec:notation}.

The superconformal algebra is the same as (\ref{eq:SCalgebra}) given
in component approach, if we perform a suitable translation of
generators between two approaches (see Table~(\ref{eq:SG})). Here we
refer to only a few characteristic commutation relations
\begin{equation}
\begin{split}
& \{Q_\alpha,\b Q_{\d\alpha}\}=-2i(\sigma^a)_{\alpha\d\alpha}P_a, \qquad
\{S_\alpha,\b S_{\d\alpha}\}=2i(\sigma^a)_{\alpha\d\alpha}K_a, \\
&[S_\alpha,P_a]=i(\sigma_a)_{\alpha\d\beta}\b Q^{\d\beta}, \qquad
\{S_\alpha,Q_\beta\}=(2D-3iA)\epsilon_{\alpha\beta}-2M_{\alpha\beta}, \\
&[\b S^{\d\alpha},P_a]=i(\b\sigma_a)^{\d\alpha \beta}Q_\beta, \qquad 
\{\b{S}^{\d\alpha},\b{Q}^{\d\beta}\} =(2D+3iA)\epsilon^{\d\alpha \d\beta}
-2M^{\d\alpha\d\beta}, \\[1mm]
&\hbox{with}\qquad 
M_{\alpha\beta}=(\sigma^{ba}\epsilon)_{\alpha\beta}M_{ab},  \qquad
M^{\d\alpha\d\beta}=(\b\sigma^{ba}\epsilon)^{\d\alpha\d\beta}M_{ab}.
\end{split}
\end{equation}
Note that the normalizations of $Q$, $S$, $A$ are different from the
component approach. The gauge superfields corresponding to the
superconformal group are denoted as 
\begin{equation}
{h_M}^\cal{A}X_{\cal{A}}={E_M}^AP_A+\df{1}{2}{\phi_M}^{ba}M_{ab}
+B_M D+A_M A +{f_M}^A K_A,
\end{equation}
where we use the calligraphic index $\cal{A}$ for the total
superconformal algebra, while the Roman uppercase index $A$ for the
set of Lorentz vector and spinor as
$P_A=(P_a,Q_\alpha,\bar{Q}^{\d\alpha})$ and 
$K_A=(K_a,S_\alpha,\b S ^{\d\alpha})$, and the index $M$ is the set
of curved indices, for example, $A_M=(A_m,A_\mu,A^{\d\mu})$. We assume
that the vierbein ${E_M}^A$ is invertible:
\begin{equation}
{E_M}^A{E_A}^N = {\delta_M}^N, \qquad {E_A}^M{E_M}^B={\delta_A}^B.
\end{equation}
The gauged superconformal transformations are taken by real parameter
superfields. These parameter superfields are denoted as
\begin{equation}
\xi^{\cal{A}}X_{\cal{A}}
=\xi(P)^AP_A+\df{1}{2}\xi(M)^{ab}M_{ba}+\xi(D) D+\xi(A)A+{\xi}(K)^{A}K_A.
\end{equation}
The gauge fields receive the superconformal transformation 
$\delta_G(\xi^{\cal{A}'}X_{\cal{A}'})$ as
\begin{equation}
\delta_G(\xi^{\cal{B}'}X_{\cal{B}'}) {h_M}^{\cal{A}}
=\der_M \xi^{\cal{B}'}{\delta_{\cal{B}'}}^{\cal{A}}
+{h_M}^{\cal{C}}\xi^{\cal{B}'}{f_{\cal{B'C}}}^\cal{A},
\end{equation}
where the primed calligraphic index $\cal{A}'$ means all the
superconformal generators other than $P_A$, namely, 
$X_{\cal{A}}=(\,P_A,\,X_{\cal{A}'}\,)$. Note that Ref.~\cite{bib:B1}
uses the different notation that $X_{\cal{A}}$ was expressed as $X_A$
in no distinction from $A$ for $(a,\,\alpha,\,\d\alpha\,)$, and our 
$h_M{}^{\cal{A}}$ and $h_M{}^{\cal{A}'}$ were denoted by $W_M{}^A$ and 
$h_M{}^{\ul{a}}$, respectively.

In the same spirit as component approach, the $P_A$ transformation is
defined to be related to the general coordinate transformation
$\delta_{\text{GC}}$ using field-independent parameter superfield $\xi^A$ as
\begin{equation}
\delta_G(\xi^AP_A) = \delta_{\text{GC}}(\xi^M:=\xi^A{E_A}^M)
-\delta_G(\xi^M{h_M}^{\cal{B}'}X_{\cal{B}'}),
\label{eq:P_Atrf}
\end{equation}
where $\xi(P)^A$ is abbreviated to $\xi^A$.
The $P_A$ transformation acting on a superfield $\Phi$ with no curved
index defines the covariant derivative as
\begin{equation}
\delta_G(\xi^AP_A)\Phi=\xi^AP_A\Phi = \xi^M\na_M\Phi 
=\xi^M(\der_M-{h_M}^{\cal{A}'}X_{\cal{A}'})\Phi.
\end{equation}
That is, $P_A=\na_A=E_A{}^M\na_M$ on superfields with flat indices. The
curvature $R_{MN}{}^{\cal{A}}$ is defined as
\begin{equation}
{R_{MN}}^{\cal{A}} = \der_M{h_N}^{\cal{A}}-\der_N{h_M}^{\cal{A}}
-({E_N}^{C}{h_M}^{\cal{B}'}-{E_M}^{C}{h_N}^{\cal{B}'})
{f_{\cal{B}'C}}^{\cal{A}} -
{h_N}^{\cal{C}'}{h_M}^{\cal{B}'}{f_{\cal{B'C'}}}^{\cal{A}}.
\label{eq:SScurvature}
\end{equation}
Here and hereafter, we use the convention of ``implicit grading''. In
superspace, we generally treat both bosonic and fermionic quantities
at the same time by the index $A$ or $M$, and should be careful for
grading of fermionic objects such as 
$X_{AB}=(-)^{a(b+n)}{E_B}^N{E_A}^M X_{MN}$,\ 
$[\na_A,\na_B\}=\na_A\na_B-(-)^{ab}\na_B\na_A$, and $Z=(-)^a{Y_A}^A$.
The grading is uniquely determined if the standard order of indices is
specified. For example, the standard order of $X_{AB}$ is $AB$ and
hence ${E_B}^N{E_A}^M X_{MN}$ should be accompanied by the grading
factor $(-)^{a(b+n)}$ since one jumps the index $A$ over two indices
$B$ and $N$ of ${E_B}^N$ in order to recover the standard order
$AB$. The implicit grading means the understanding of omitting such
unique grading factors from everywhere. In other words, we can treat
the indices $A$, $M$ as if they were bosonic ones. The same implicit
grading convention is used also for the index $\cal{A}$ of 
superconformal generators. In the definition of curvatures, the
commutation relation of $P_A$ is as follows
\begin{equation}
\begin{split}
& [P_A,P_B] = -{R_{AB}}^{\cal{C}}X_{\cal{C}}
=-R(P)_{AB}{}^CP_C-\df{1}{2}R(M)_{AB}{}^{dc}M_{cd}  \\
& \hspace{13em} -R(D)_{AB}D-R(A)_{AB}A-R(K)_{AB}{}^CK_C,
\end{split}
\end{equation}
where $R_{AB}{}^{\cal{C}}={E_B}^N{E_A}^M R_{MN}{}^{\cal{C}}$ in terms
of ${R_{MN}}^\cal{C}$ given in (\ref{eq:SScurvature}). In
Ref.~\cite{bib:B1}, $R(P)_{AB}{}^C$ is expressed as $T_{AB}{}^C$,
$R(M)_{AB}{}^{cd}$ is $R_{AB}{}^{cd}$, $R(D)_{AB}$ is $H_{AB}$, and
$R(A)_{AB}$ is $F_{AB}$.

Several constraints are imposed on the curvature superfields to
eliminate the redundant degrees of freedom. First, the constraints on
$R_{\ul{\alpha}\ul{\beta}}$ are as follows
\begin{equation}
\begin{split}
& {R_{\alpha\beta}}^{\cal{A}} = 0, \qquad
{R_{\d\alpha\d\beta}}^{\cal{A}} = 0, \qquad
{R(P)_{\alpha\d\beta}}^{c} = 2i(\sigma^c)_{\alpha\d\beta}, \\[1mm]
& \hspace*{20mm} 
{R_{\alpha\d\beta}}^{\cal{A}} = 0 \quad (\text{otherwise}),
\end{split}
\label{eq:constraintI}
\end{equation}
which guarantees the commutation relations of covariant spinor
derivatives to take the simple form (as in the global supersymmetry case)
\begin{equation}
\{\na_\alpha, \na_\beta \}=0, \qquad 
\{\b\na_{\d\alpha},\b\na_{\d\beta}\}=0, \qquad
\{\na_\alpha,\b\na_{\d\beta}\}=-2i\na_{\alpha\d\beta}.
\label{eq:nablaAlgebra}
\end{equation}
Secondly, the following constraints on $R_{\ul{\alpha} a}$ are imposed
\begin{equation}
{R(P)_{\ul{\gamma} b}}^A=0, \qquad
R(D)_{\ul{\beta} a}=0, \qquad
R(A)_{\ul{\beta} a}=0.
\label{eq:constraintII}
\end{equation}
By solving the Bianchi identities
\begin{equation}
 [\na_A,[\na_B,\na_C]]+
[\na_B,[\na_C,\na_A]]+[\na_C,[\na_A,\na_B]]=0
\end{equation}
under these constraints (with implicit grading understood), 
one finds that all other non-vanishing curvatures can be expressed by
a single superfield $W_{\alpha\beta\gamma}$ with totally-symmetric
undotted spinor indices $\alpha$, $\beta$, $\gamma$ as seen below.

The Bianchi identities with the first constraints
(\ref{eq:constraintI}) imply that the curvatures $R_{\ul{\alpha} b}$
and $R_{ab}$ can be expressed by a ``gaugino'' superfield 
$\cal{W}_{\ul{\alpha}}$, which is superconformal algebra valued,
\begin{gather}
R_{\alpha,\beta\d\gamma} = -[\na_\alpha,\na_{\beta\d\gamma}]
=2i\epsilon_{\alpha\beta}\cal{W}_{\d\gamma},  \qquad
R_{\d\alpha,\d\beta\gamma} = -[\b\na_{\d\alpha},\na_{\d\beta\gamma}]
=2i\epsilon_{\d\alpha\d\beta}\cal{W}_\gamma, 
\label{eq:Rspinorvector}  \\[1mm]
R_{\alpha\d\alpha,\beta\d\beta} = -\epsilon_{\d\alpha\d\beta}
\{\na_{(\alpha},\cal{W}_{\beta)}\} -\epsilon_{\alpha\beta}
\{\b\na_{(\d\alpha},\cal{W}_{\d\beta)}\},
\label{eq:RvvW}
\end{gather}
where $R_{\alpha,\beta\d\gamma}=(\sigma^b)_{\beta\d\gamma}R_{\alpha b}$,
which is $R_{\alpha (\beta\d\gamma)}$ in Ref.~\cite{bib:B1}.
The brackets $(\ )$ on the indices imply the symmetrization 
with weight one, e.g.,
$\psi_{(\alpha}\chi_{\beta)}=(1/2)(\psi_{\alpha}\chi_{\beta}
+\psi_{\beta}\chi_{\alpha})$.
This algebra-valued superfield $\cal{W}_{\ul{\alpha}}$ satisfies 
\begin{gather}
\{\na_\alpha,\cal{W}_{\d\gamma}\}=
\{\b\na_{\d\alpha},\cal{W}_\gamma\}=0, \quad \text{(chirality)}
\label{eq:chiral}  \\[1mm]
\{\na^\alpha,\cal{W}_\alpha\}=
\{\b\na_{\d\beta},\cal{W}^{\d\beta}\}, \qquad \text{(reality)} 
\label{eq:real}  \\ \quad
[M_{bc},\cal{W}_\alpha]={(\sigma_{bc})_\alpha}^\beta\cal{W}_\beta, \quad
[D,\cal{W}_\alpha]=\df{3}{2}\cal{W}_\alpha, \quad\;
[A,\cal{W}_\alpha]=i\cal{W}_\alpha, \quad\;
[K_A,\cal{W}_\alpha]=0.
\end{gather}
The further input of the second constraints (\ref{eq:constraintII}) 
implies that $\cal{W}_{\ul{\alpha}}$ has no $P_A$, $D$, $A$ components,
$\cal{W}(P)_{\ul{\alpha}}{}^A
=\cal{W}(D)_{\ul{\alpha}}=\cal{W}(A)_{\ul{\alpha}}=0$, so that 
\begin{equation}
\cal{W}_\alpha = \df{1}{2}{\cal{W}(M)_\alpha}^{bc}M_{cb}
+{\cal{W}(K)_\alpha}^BK_B.
\label{eq:gaugino}
\end{equation}
With the help of the superconformal algebra, the chirality and reality
conditions (\ref{eq:chiral}) and (\ref{eq:real}) leads to the final
expression
\begin{align}
\cal{W}_\alpha &=
(\epsilon\sigma^{bc})^{\beta\gamma}W_{\alpha\beta\gamma}M_{cb}
+\df{1}{2}\(\na^{\gamma}{W_{\gamma\alpha}}^\beta\)S_\beta
-\df{1}{2}\(\na^{\gamma\d\beta}{W_{\gamma\alpha}}^\beta\)K_{\beta\d\beta}, \\
{\cal{W}}^{\d\alpha} &=
(\b\sigma^{bc}\epsilon)^{\d\gamma\d\beta}
W^{\d\alpha}{}_{\d\beta\d\gamma}M_{cb}
-\df{1}{2}\(\b\na_{\d\gamma}{W}^{\d\gamma\d\alpha}_{
\hphantom{\d\gamma\d\alpha}\d\beta} \)\b{S}^{\d\beta}
-\df{1}{2}\( \na^{\d\gamma\beta}W_{\d\gamma}^{
\hphantom{{\d\gamma}}\d\alpha\d\beta} \)K_{\beta\d\beta}.
\end{align}
In this way, the gaugino superfield $\cal{W}_{\alpha}$ is expressed by
the totally symmetric superfield $W_{\alpha\beta\gamma}$ which satisfies
\begin{equation}
\b\na_{\d\alpha}W_{\alpha\beta\gamma}=0, \quad\;
D W_{\alpha\beta\gamma}=\df{3}{2}W_{\alpha\beta\gamma}, \quad\;
AW_{\alpha\beta\gamma}=iW_{\alpha\beta\gamma}, \quad\;
K_AW_{\alpha\beta\gamma}=0.
\label{eq:SCW}
\end{equation}
Owing to Eqs.~(\ref{eq:Rspinorvector}) and (\ref{eq:RvvW}), all the
curvatures $R_{AB}$ can also be written in terms of
$W_{\alpha\beta\gamma}$, its conjugate, and their covariant
derivatives. In particular, the $R_{ab}$ component is expressed as
\begin{equation}
\begin{split}
R_{\alpha\d\alpha,\beta\d\beta} = &\,
\epsilon_{\d\alpha\d\beta}\(\,
2{W_{\alpha\beta}}^\gamma Q_\gamma
+\na^\gamma{W^\delta}_{\alpha\beta}M_{\delta\gamma}
+\na^\gamma W_{\gamma\alpha\beta}D
-\df{3}{2}i\na^\gamma W_{\gamma\alpha\beta}A
\right. \\
&\left. \hphantom{\epsilon_{\d\alpha\d\beta}\big(}
+\df{1}{4}\na^2 W_{\alpha\beta}^{\hphantom{\alpha\beta}\gamma}S_\gamma
-i\na^\gamma_{\hphantom{\gamma}\d\gamma}W_{\gamma\alpha\beta}\b{S}^{\d\gamma}
+\df{1}{2}\na_\alpha\na^{\d\beta\gamma }
W_{\gamma\beta}^{\hphantom{\gamma\beta}\delta}K_{\delta\d\beta} \)  \\
& +\epsilon_{\alpha\beta}\(
-2W_{\d\alpha\d\beta\d\gamma}\b{Q}^{\d\gamma}
+\bar\na_{\d\gamma}W_{\d\delta\d\alpha\d\beta}M^{\d\gamma\d\delta}
+\bar\na_{\d\gamma}W^{\d\gamma}{}_{\d\alpha\d\beta}D
+\df{3}{2}i\b\na_{\d\gamma}W^{\d\gamma}{}_{\d\alpha\d\beta}A
\right. \\
&\left. \hphantom{+\epsilon_{\alpha\beta}\big(}
-\df{1}{4}\b\na^2 W_{\d\beta\d\alpha\d\delta}\b{S}^{\d\delta}
-i\na^{\d\gamma\gamma}W_{\d\gamma\d\alpha\d\beta}S_\gamma
+\df{1}{2}\b\na_{\d\alpha}\na^{\d\gamma\beta}
W_{\d\gamma\d\beta}{}^{\d\delta}K_{\beta\d\delta} \).
\end{split}
\label{eq:Rssss}
\end{equation}

Now the concept of {\it primary} superfield is introduced to describe 
matter superfields, invariant action over the superspace, and so on. A
primary superfield $\Phi_\Gamma$ is defined as the superfield on which
the action of superconformal group is
\begin{equation}
M_{bc}\Phi_\Gamma =(\cal{S}_{bc})_\Gamma{}^\Sigma \Phi_\Sigma, \quad\;
D\Phi_\Gamma=\Delta \Phi_\Gamma, \quad\;
A\Phi_\Gamma=iw\Phi_\Gamma, \quad\;
K_A\Phi_\Gamma=0.
\label{eq:Bpsf}
\end{equation}
where $\Gamma$ and $\Sigma$ represent general Lorentz indices such as
$\Gamma=(\alpha_1,\ldots,\alpha_n,\d\beta_1,\ldots,\d\beta_m)$, and 
$\cal{S}_{bc}$ is the representation matrix of Lorentz algebra which
$\Phi_\Gamma$ belongs to. The real constant numbers $\Delta$ and $w$
are called the Weyl and chiral weights, respectively. 
The last property $K_A\Phi_\Gamma=0$ is most important for
$\Phi_\Gamma$ being primary. That is generally violated for its
derivative $\na_A\Phi_\Gamma$. As for $W_{\alpha\beta\gamma}$, 
Eqs.~(\ref{eq:SCW}) imply it is a primary chiral superfield with
Weyl weight $\Delta=3/2$ and chiral weight $w=1$, where a chiral
superfield means that it satisfies $\b\na^{\d\alpha}\Phi=0$ as
usual. It should be noted that this chirality condition is
superconformally covariant.
 
An invariant integral over the superspace is given by
\begin{equation}
 S_D = \int d^4xd^4\theta\, E V,
\end{equation}
where $E=\det({E_M}^A)$. Here we are using implicit grading and
omitting to write the superdeterminant ``sdet''. The superconformal
transformation law of the density $E$ is
\begin{gather}
\delta_G(\xi^{\cal{A}'}X_{\cal{A}'})E
= E{E_A}^M \delta_G(\xi^{\cal{A}'}X_{\cal{A}'})E_M{}^A
= E{E_A}^M {h_M}^{\cal{C}}\xi^{\cal{B}'}{f_{\cal{B}'\cal{C}}}^A
= E\xi^{\cal{B}'}{f_{\cal{B}'A}}^A  \nonumber \\[1mm]
\to \quad
D\,E = -2\,E, \quad\; M_{ab}\,E=A\,E=K_A\,E =0, \quad
\label{eq:E}
\end{gather}
since the superconformal generators $X_{\cal{B}'}$ other than $P_A$ 
carry non-positive Weyl weights so that the commutator 
$[X_{\cal{B}'},\,X_{\cal{C}}]$ yields positive Weyl weight $P_A$ 
only when $X_{\cal{C}}=P_C$ (and $X_{\cal{B}'}=D$ or $A$), in which
case ${E_A}^M {h_M}^{\cal{C}}=
{E_A}^M {E_M}^{C}= \delta_A{}^C$. From (\ref{eq:E}), the invariance
conditions for the action $S_D$ become
\begin{equation}
D\,V = 2\,V,\quad\; A\,V= M_{ab}\,V=K_A\,V =0.
\end{equation}
That is, $V$ must be a $(\Delta,w)=(2,0)$ primary real superfield with
no Lorentz index. The invariance of $S_D$ under the GC transformation
in superspace is manifest and hence invariant under the $P_A$
transformation. Thus the action $S_D$ is fully superconformal
invariant, called the D-type integration.

The superconformal counterpart of the $d^2\theta$ integral in global
supersymmetry is 
\begin{equation}
S_F = \int d^4x d^2\theta \,\cal{E} W.
\label{eq:S_F}
\end{equation}
The chiral density $\cal{E}$ is given by the superdeterminant of
vielbein in the chiral subspace with dotted spinor directions being 
omitted from ${E_M}^A$, that is, $\cal{E}=
\det{\cal{E}_{\mathfrak{m}}{}^{\mathfrak{a}}}$ 
with $\cal{E}_{\mathfrak{m}}{}^{\mathfrak{a}}=
E_{\mathfrak{m}}{}^{\mathfrak{a}}$, $\mathfrak{a}=(a,\alpha)$, 
and $\mathfrak{m}=(m,\mu)$. In (\ref{eq:S_F}), $W$ is a covariantly
chiral superfield defined by $\bar\na_{\d\alpha}W=0$. The invariance
of the action $S_F$ requires that $W$ must be a $(\Delta,w)=(3,2)$
primary chiral superfield with no Lorentz index. Since the integral
$S_F$ does not depend on $\b\theta$, it is supposed to be executed at
$\b\theta=0$, which is called the F-type integration. Performing the
$d^2\theta$ integration in (\ref{eq:S_F}), we obtain the component
expression of the F-type integration as 
\begin{equation}
\int d^4x d^2\theta \,\cal{E}W =\int d^4x\,e \(-\df{1}{4}\na^2 W
+\df{i}{2}\b\psi_{a\d\alpha}(\b\sigma^a)^{\d\alpha\beta}\na_\beta W
-\(\b\psi_a\b\sigma^{ab}\b\psi_{b}\)W \)_{\!\theta=\b\theta=0}. 
\label{eq:F-int}
\end{equation}
The D-type integration is related to the F-type one as
\begin{equation}
\int d^4xd^4\theta\, EV
=\df{1}{2}\int d^4x d^2\theta\,\cal{E} \cal{P}[V] 
+\df{1}{2}\int d^4x d^2\b\theta\,\b{\cal{E}} \b{\cal{P}}[V],
\label{eq:DtoF}
\end{equation}
where 
\begin{equation}
\cal{P}[V] = -\df{1}{4}\b\na^2 V
\end{equation}
is the chiral projection operator. The component expression of the
D-type integration is obtained using the equation (\ref{eq:DtoF}).

The action of the matter coupled SUGRA system is given in conformal
superspace as
\begin{equation}
 S=-3\int d^4x d^4\theta\, E\,\Phi^\co\b\Phi^\co e^{-K/3}
+\(\int d^4x d^2 \theta\, \cal{E}\,(\Phi^\co)^3W +\text{h.c.}\),
\end{equation}
where $\Phi^\co$ is the compensator chiral superfield carrying Weyl and
chiral weights $(\Delta,w)=(1,2/3)$. The K\"ahler potential $K$ and 
the superpotential $W$ are the functions of chiral matter superfields
$\Phi^i$ with weights $(\Delta,w)=(0,0)$. In addition, $K$ is a real
function and $W$ is holomorphic. The gauge-fixing conditions leading to
Poincar\'e SUGRA with the canonically normalized EH term are given in
Ref.~\cite{bib:B1}:
\begin{equation}
D,\ A\hbox{-gauge}:\ \Phi^\co=\b\Phi^\co=e^{K/6}, \qquad 
K_A\hbox{-gauge}:\ B_M=0.
\label{eq:gfconf}
\end{equation}
From the gauge-fixing condition for $K_A$-gauge, the $D$ gauge field
$B_M$ vanishes and the $K_A$ gauge field ${f_M}^A$ loses the gauge
freedom. So ${f_M}^A$ drops out from the covariant derivative and
curvatures. The derivative after gauge fixing is written as
$\cal{D}_A=\na_A+{f_A}^BK_B$, and the curvatures after gauge fixing 
are written in terms of conformal curvatures and the gauge-fixed
${f_M}^A$. Moreover, the constraints for the conformal curvatures give 
the constraint for the gauge-fixed ${f_M}^A$. The constraints for
$R(D)_{\ul{\alpha}\ul{\beta}}$ become the constraints for 
$f_{\ul{\alpha}}{}^{\ul{\beta}}$. The curvature $R(D)_{AB}$ is written as
\begin{equation}
R(D)_{AB}=E_A{}^ME_B{}^N (\der_MB_N-\der_N B_M)+2f_{AB}(-)^a-2f_{BA}(-)^b.
\end{equation}
The constraints for $R(D)_{\ul{\alpha}\ul{\beta}}$ lead
\begin{equation}
f_{\alpha\beta}=-\epsilon_{\alpha\beta}\b{R}, \qquad
f_{\d\alpha\d\beta}=\epsilon_{\d\alpha\d\beta}R, \qquad
f_{\alpha\d\beta}=-f_{\d\beta\alpha}=-\df{1}{2}G_{\alpha\d\beta}. 
\end{equation}
The constraint $R(D)_{\ul{\alpha} b}=0$ implies
\begin{equation}
f_{\ul{\alpha}b}=-f_{b\ul{\alpha}}.
\end{equation}
The constraints $R(K)_{\alpha\beta,\d\gamma}=0$ and
$R(K)_{\alpha\d\gamma}{}^\beta=0$ and their conjugates give 
\begin{align}
3if_{\alpha,\beta\d\beta} &=
\df{1}{2}\cal{D}_\alpha G_{\beta\d\beta}
+\cal{D}_\beta G_{\alpha\d\beta}+\epsilon_{\alpha\beta}
\b{\cal{D}}_{\d\beta }\b{R}, \\
-3if_{\d\alpha,\beta\d\beta} &=
\df{1}{2}\b{\cal{D}}_{\d\alpha}G_{\beta\d\beta}
+\b{\cal{D}}_{\d\beta}G_{\beta\d\alpha}
+\epsilon_{\d\alpha\d\beta}\cal{D}_{\alpha} R.
\end{align}
Finally the constraint $R(K)_{\alpha\d\beta}{}^c=0$ means
\begin{equation}
f_{\alpha\d\alpha,\beta\d\beta} = 
\df{i}{2}(\cal{D}_\alpha f_{\d\alpha,\beta\d\beta}+
\b{\cal{D}}_{\d\alpha} f_{\alpha,\beta\d\beta})
+2\epsilon_{\alpha\beta}\epsilon_{\d\alpha\d\beta}R\b{R}
+\df{1}{2}G_{\beta\d\alpha}G_{\alpha\d\beta}.
\end{equation}
Since the conformal curvatures are written in terms of 
$W_{\alpha\beta\gamma}$, the curvatures after gauge fixing are written
in terms of $R$, $G_{\alpha\d\beta}$, $W_{\alpha\beta\gamma}$ and the
derivative $\cal{D}_A$.

It is noted that the gauge-fixing conditions (\ref{eq:gfconf}) also
fix the $A$ gauge superfield $A_M$. The covariantly chiral condition
of $\Phi^\co$ is $0=\bar\na^{\d\alpha} \Phi^\co=
E^{\d\alpha M}\der_M\Phi^\co-B^{\d\alpha}\Phi^\co-
\frac{2}{3}iA^{\d\alpha}\Phi^\co$, and further imposing the gauge
conditions $\Phi^\co=e^{K/6}$ and $B^{\d\alpha}=0$ leads to
\begin{equation}
A^{\d\alpha}=-\df{i}{4}K_{i^*}{\b{\cal{D}}}^{\d\alpha}\b\Phi^{i^*},
\end{equation}
where ${\b{\cal{D}}}^{\d\alpha}\b\Phi^{i^*}=
E^{\d\alpha M}\der_M\b\Phi^{i^*}$
and $K_{i^*}={\der K}/{\der\b\Phi^{i^*}}$. The chirality condition for
matter superfields $\Phi^i$ are used; $0=\b\na^{\d\alpha} \Phi^i
=E^{\d\alpha M}\der_M\Phi^i={\b{\cal{D}}}^{\d\alpha}\Phi^i$. In the
same way, from $\na_\alpha \b\Phi^\co =0$, $A_\alpha$ is fixed as 
\begin{equation}
A_\alpha =\df{i}{4}K_i \cal{D}_\alpha \Phi^i.
\end{equation}
Similarly, from the relation $\b\na^{\d\alpha}\na_\alpha\Phi^\co
=-2i{\na_\alpha}^{\d\alpha}\Phi^\co$, we obtain
\begin{equation}
{A_\alpha}^{\d\alpha} = \df{i}{4}(K_i{\cal{D}_\alpha}^{\d\alpha}\Phi^i
-K_{i^*}{\cal{D}_\alpha}^{\d\alpha}\b\Phi^{i^*} )
-\df{3}{2}{G_\alpha}^{\d\alpha} +\df{1}{4}g_{ij^*}(\cal{D}_\alpha
\Phi^i)({\b{\cal{D}}}^{\d\alpha}\b\Phi^{j^*} ),
\end{equation}
where ${G_{\alpha}}^{\d\alpha}=-2f_\alpha{}^{\d\alpha}$ and 
$g_{ij^*}=\der^2K/\der\Phi^i\der\Phi^{j^*}$.

\section{Correspondence between component and superspace approaches}
\label{sec:corr}

In this section we present the correspondence between component and
superspace formulations. The objects which we deal with are the
superconformal algebra, gauge fields, curvatures and their
constraints, conformal multiplets with external Lorentz indices,
chiral projection, and invariant actions. Note that the notations and
conventions are different in two approaches and the dictionary between
them is given in appendix \ref{sec:notation} for spinors, vectors,
gamma matrices, and tensors.

\subsection{Superconformal algebra, gauge fields and curvatures}

As discussed in Sec.~\ref{sec:comapp}, the $Q$ and $P_a$
transformation in component approach are deformed from the original
group laws. In the following, we use only the final form of them and
the deformed $\tilde P_a$ transformation is simply denoted as $P_a$.

Let us begin with the dictionary for the normalization of 
superconformal generators and Weyl and chiral weights. The
correspondence is given by
\begin{equation}
\begin{array}{c|c}
\hline\hline
\text{component} & \text{superspace} \\ \hline
P_a,\;\; 2Q_\alpha,\;\; \Tspan{2\bar{Q}^{\d\alpha}}
& (P_a,\;Q_\alpha,\;\bar{Q}^{\d\alpha})=P_A
\\ \hline 
-M_{ab},\;\; D,\;\; \Tspan{\tfrac43A}
& M_{ab},\;\; D,\;\; A 
\\ \hline
K_a,\; -2S_\alpha,\; \Tspan{-2\bar{S}^{\d\alpha}}
& (K_a,\; S_\alpha,\; \bar{S}^{\d\alpha})= K_A
\\ \hline
w,\;\; \Tspan{\tfrac23n}
& {\Delta},\;\; w
\\ \hline
\end{array}
\label{eq:SG}
\end{equation}
The correspondence of gauge parameters is set to satisfy
$\epsilon^AX_A \corr \xi^{\cal{A}}|X_{\cal{A}}$ and given by
\begin{equation}
\begin{array}{c|c|c}
\hline\hline
& \text{component} & \text{superspace} \\ \hline
P,\;\; Q & \Tspan{(\xi^a,\ \tfrac12\b\varepsilon)} & \;
\Tspan{(\xi(P)^a,\; \xi(P)^{\alpha},\; \b\xi(P)_{\d\alpha})|\,=\,\xi(P)^A|}
\\ \hline
M,\; D,\; A & \lambda^{ab},\ \rho,\;\, \Tspan{\frac{3}{4}}\theta  & 
\xi(M)^{ab}|,\ \ \xi(D)|,\ \ \xi(A)| \\ \hline
K,\;\; S & \Tspan{(\xi_K^a,\ \tfrac{-1}{2}\b\zeta)} & \;
\Tspan{(\xi(K)^a,\,\xi(K)^\alpha,\,\b\xi(K)_{\d\alpha})|\,=\,\xi(K)^A|} 
\\ \hline
\end{array}
\label{eq:parameter}
\end{equation}
The vertical bar ``$|$'' means the $\theta=\b\theta=0$ projection,
i.e., the lowest component of superfield. Since gauge fields $\times$ 
generators essentially represent the common quantity in both approaches, 
${h_\mu}^{A}X_{A} \corr {h_m}^{\cal{A}}|X_{\cal{A}}$, the
correspondence of gauge fields appears with inverse normalizations of
the generators
\begin{equation}
\begin{array}{c|c|c}
\hline\hline
& \text{component} & \text{superspace} \\ \hline
P,\;\; Q & ({e_\mu}^a,\ \tfrac12\psi_\mu) & 
\Tspan{E_m{}^A| =(E_m{}^a,\,E_{m\alpha},\,{E_m}^{\d\alpha})|= (e_m{}^a,\, 
\tfrac12\psi_{m\alpha},\, \tfrac12{\b{\psi}_m}{}^{\d\alpha})} 
\\ \hline
M,\; D,\; A & {\omega_\mu}^{ab},\ b_\mu,\ 
\Tspan{\frac{3}{4}} A_\mu & 
{\phi_m}^{ab}| = {\omega_m}^{ab},\ \ B_m|,\ \ A_m| 
\\ \hline
K,\;\; S & ({f_\mu}^a,\,-\tfrac12\varphi_\mu) & f_m{}^A| =  
\Tspan{(f_m{}^a,\,f_{m\alpha},\,{f_m}^{\d\alpha})}|
\\ \hline
\end{array}
\label{eq:GF}
\end{equation}
In the table, the curved index $\mu$ of component approach corresponds
to the index $m$ of superspace.

The curvature in superspace, ${R_{mn}}^{\cal{C}}$, with curved tensor
indices was defined in Eq.~(\ref{eq:SScurvature}). The lowest
component of flat indexed curvature superfield $R_{ab}{}^{\cal{C}}$ is 
given by
\begin{equation}
\begin{split} \quad
{R_{ab}}^{\cal{C}}| = {E_b}^N{E_a}^M{R_{MN}}^{\cal{C}}|
& = {e_a}^m{e_b}^n {R_{mn}}^{\cal{C}}| 
-\df{i}{2}\({\psi_a}^\alpha(\sigma_b)_{\alpha\d\beta}-{\psi_b}^\alpha 
(\sigma_a)_{\alpha\d\beta} \)\cal{W}^{\d\beta\cal{C}}|  \\
&\hspace{2em}
+\df{i}{2}\({\b\psi_{a\d\alpha}}(\b\sigma_b)^{\d\alpha\beta}
-\b\psi_{b\d\alpha}(\b\sigma_a)^{\d\alpha\beta}\)
\cal{W}_\beta{}^{\cal{C}}| 
+\df{1}{4}{\psi_{a}}^{\ul{\alpha}}
{\psi_{b}}^{\ul{\beta}}{R_{\ul{\alpha}\ul{\beta}}}^{\cal{C}}|. 
\end{split}
\end{equation}
Using the correspondence of gauge fields given in (\ref{eq:GF}), we
find that the curvatures coincide with (the negative of) the covariant
curvatures with the algebra deformation of component approach, up to
the normalization of generators
\begin{equation}
\begin{array}{c|c}  \hline\hline
  \text{component} & \text{superspace}  \\ \hline 
-\bigl(\,R_{ab}(P^c),\;\; \tfrac12R_{ab}(Q)\,\bigr) 
& \Tspan{\bigl(\,{R(P)_{ab}}^c,\ {R(P)_{ab}}^\gamma,\ 
R(P)_{ab\d\gamma}\,\bigr)\big|={R(P)_{ab}}^C\big|} 
\\ \hline
-R_{ab}^{\text{cov}}(M^{cd}),\ -R_{ab}(D),\;\;
\Tspan{\frac{-3}{4}}R_{ab}(A) 
& {R(M)_{ab}}^{cd}|,\ \ R(D)_{ab}|,\ \ R(A)_{ab}|
\\ \hline
-\Tspan{\big(\,R_{ab}^{\text{cov}}(K^c),\ 
-\tfrac12R_{ab}^{\text{cov}}(S)\,\bigr)}
& \Tspan{\bigl(\,{R(K)_{ab}}^c,\ {R(K)_{ab}}^\gamma,\ 
R(K)_{ab\d\gamma}\,\bigr)\big|={R(K)_{ab}}^C\big|} \\
\hline
\end{array}
\label{eq:CurvatureCorr}
\end{equation}
The `covariantization' is necessary only for the $M$, $S$, $K_a$
curvatures in component approach, which correspond in superspace to
the fact that the gaugino superfield $\cal{W}_{\ul{\alpha}}$ has
non-vanishing components only for the $M$, $S$, $K_a$ generators. In
obtaining the correspondence table (\ref{eq:CurvatureCorr}), we have
used the following relations of $\cal{W}_\alpha$ to the curvatures in
component approach
\begin{equation}
\begin{array}{c|l} \hline\hline
\text{component} & \multicolumn{1}{c}{\text{superspace}} 
\\ \hline
R_{ab}(Q)\gamma_5 &
\Tspan{\bmx \cal{W}(M)^\alpha{}_{ab},& \cal{W}(M)_{\d\alpha,{ab}} \emx\!|
=-2 \bmx R(P)_{ab}{}^\alpha, & -R(P)_{ab,\d\alpha} \emx\!|}
\\ \hline
-\df{i}{4}\sigma^{ab}\gamma_5 R_{ab}(A) &
\Tspan{\bmx \cal{W}(K)_\alpha{}^{\beta} & 0 \\
0 & \cal{W}(K)^{\d\alpha}{}_{\d\beta} \emx\!\Big|} \\
& \ \ = \Tspan{-\df{i}{3} \bmx (\sigma^{ab})_\alpha{}^\beta & 0 \\
0 & (\b\sigma^{ab})^{\d\alpha}{}_{\d\beta} \emx}
\axg R(A)_{ab}| 
\\ \hline
-\df{1}{4} R^{\text{cov}}_{bc}(S)\gamma^c\gamma_5
& \bmx \cal{W}(K)^{\alpha}{}_{b} & \cal{W}(K)_{\d\alpha} {}_b \emx \!| \\
\ \ +\df{1}{4} \tilde{R}^{\text{cov}}_{bc}(S)\gamma^c
& \ \ = -\df{i}{2} \bmx R(K)_{bc}{}^{\beta} & R(K)_{bc,\d\beta} \emx \!|
\stog{}{c}{\beta}{\alpha}\axg  \\
& \qquad
+\df{1}{2} \bmx (*R(K))_{bc}{}^{\beta} & (*R(K))_{bc,\d\beta} \emx \!|
\stog{}{c}{\beta}{\alpha}
\\ \hline
\end{array} 
\end{equation}
Note that these quantities stand for the spinor-vector components of  
superspace curvature $R_{\ul{\alpha} b}{}^{\cal{C}}$ because of the 
relations (\ref{eq:Rspinorvector}). 

The correspondences of the curvatures (\ref{eq:CurvatureCorr}) are
summarized in a simple expression
\begin{equation}
{R^{\text{cov}\,C}_{ab}} X_C \;\; \text{(component)} \quad
\corr \quad -{R_{ab}}^{\cal{C}}|\, X_{\cal{C}} \;\; \text{(superspace)}.
\label{eq:corrRab}
\end{equation}
We emphasize that such identification holds for the 
{\it flat indexed curvatures}, while it does for the 
{\it curved indexed gauge fields}
\begin{equation}
{h_{\mu}}^C X_{C} \;\; \hbox{(component)} \quad
\corr \quad  {h_{m}}^{\cal{C}}|\, X_{\cal{C}} \;\; \hbox{(superspace)}.
\label{eq:corrgaugefield}
\end{equation}
For instance, the component approach counterpart of the flat indexed 
gauge field in superspace is found through the expression
\begin{equation}
h_a{}^{\cal{C}}| = {E_a}^Mh_M{}^{\cal{C}}| = e_a{}^m h_m{}^{\cal{C}}| 
-\df12 {\psi_{a}}^{\ul{\alpha}} h_{\ul{\alpha}}{}^{\cal{C}}|. 
\end{equation}

The constraints on curvatures also have the correspondence, though the
constraints in superspace are directly imposed on the spinor-spinor or
spinor-vector component of curvatures. The restricted form of the
vector-vector component $R_{ab}$ in superspace is derived from other
constraints and explicitly written in terms of the primary chiral
superfield $W_{\alpha\beta\gamma}$. That is, Eq.~(\ref{eq:Rssss})
implies the following expressions for the curvatures
$R(X)_{ab}{}^{\cal{A}}$ in superspace
\begin{equation}
\begin{array}{c|c}
\hline \hline
\Tspan{\text{curvatures $R(X)_{ab}{}^{\cal{A}}$ in superspace}} & 
\\ \hline
\Tspan{ R(P)_{ab}{}^c } & 0
\\ \hline
\Tspan{R(P)^-_{\gamma\beta}{}^\alpha, \quad
R(P)^+_{\d\gamma\d\beta,\,\d\alpha}} 
&
W_{\gamma\beta}{}^\alpha, \quad W_{\d\gamma\d\beta\d\alpha} 
\\ \hline
\tfrac{1}{4}R(M)^{-,\,-}_{\beta\alpha,\,\delta\gamma}, \quad
\Tspan{\tfrac{1}{4}R(M)^{+,\,+}_{\d\beta\d\alpha,\,\d\delta\d\gamma}} 
&
\nabla_{(\delta} W_{\gamma)\beta\alpha}, \quad
\bar\nabla_{(\d\delta}{W}_{\d\gamma)\d\beta\d\alpha}
\\ \hline
R(D)^-_{\beta\alpha}= \tfrac23i R(A)^-_{\beta\alpha}, \quad 
R(D)^+_{\d\beta\d\alpha}=\Tspan{\tfrac{-2}{3}i R(A)^+_{\d\beta\d\alpha}}
&
-\tfrac12\na^\gamma W_{\gamma\beta\alpha}, \quad
\Tspan{\tfrac12\b\na^{\d\gamma}W_{\d\gamma\d\beta\d\alpha}}
\\ \hline
R(K)^-_{\gamma\beta}{}^\alpha, \quad
\Tspan{R(K)^+_{\d\gamma\d\beta,\,\d\alpha}}
&
\tfrac18\na^2 W_{\gamma\beta}{}^\alpha, \quad 
\tfrac18\Tspan{\b\na^2}{W}_{\d\gamma\d\beta\d\alpha}
\\ \hline
R(K)^-_{\gamma\beta}{}^a(\sigma_a)_{\alpha\d\alpha}, \quad 
R(K)^+_{\d\gamma\d\beta}{}^a(\sigma_a)_{\alpha\d\alpha} 
&
\;\Tspan{\tfrac{-1}{2}\nabla_\gamma\na^\delta{}_{\d\alpha}
W_{\delta\beta\alpha}}, \;\;
\Tspan{\tfrac12 \bar\nabla_{\d\gamma}\na_\alpha{}^{\d\delta}
W_{\d\delta\d\beta\d\alpha}} 
\\ \hline
\end{array} 
\label{eq:R-W}
\end{equation}
The chiral decomposition of anti-symmetric tensor is defined in
Eq.~(\ref{eq:chiral_decom}).

We can see the correspondence of curvature constraints using the fact
that all the curvature components $R(X)_{ab}{}^{\cal{A}}$ with
vector-vector indices are expressed by $W_{\alpha\beta\gamma}$ in
superspace. First, the constraint (\ref{eq:constP}) in component approach
is equivalent to $R_{ab}(P^c)=0$ and hence corresponds to 
$R(P)_{ab}{}^c|=0$ in superspace, as seen in Table
(\ref{eq:CurvatureCorr}) and (\ref{eq:R-W}). Secondly, the constraint
(\ref{eq:constQ}), equivalent to $R_{ab}(Q)\gamma^b=0$, corresponds to
the equation $R(P)_{ab}{}^\alpha|(\sigma^b)_{\alpha\d\delta}=0$ and its
conjugate in superspace. This is found from (\ref{eq:R-W}) that
$R(P)_{ab}{}^\alpha$ has only chiral component 
$R(P)_{\gamma\d\gamma,\beta\d\beta}{}^\alpha
=2\varepsilon_{\d\gamma\d\beta}W_{\gamma\beta}{}^\alpha$ so that
\begin{equation}
R(P)_{ab}{}^\alpha(\sigma^b)_{\alpha\d\delta}  \;\propto\;
(\b\sigma_b)^{\d\beta\beta}\varepsilon_{\d\gamma\d\beta}
W_{\gamma\beta}{}^\alpha(\sigma^b)_{\alpha\d\delta}  \;\propto\;
\varepsilon_{\d\gamma\d\delta}W_{\gamma\alpha}{}^\alpha,
\end{equation}
which vanishes since $W_{\alpha\beta\gamma}$ is a totally symmetric 
superfield. The final constraint (\ref{eq:constM}) in component
approach, which is equivalently rewritten as
\begin{equation}
R^{\text{cov}}_{ac}(M^{cb}) + \df{1}{2}i\tilde{R}^{b}{}_{a}(A)=0,
\end{equation}
corresponds to (the lowest of) the relation between $R(M)_{ab}{}^{cd}$
and $R(A)_{ab}$ as
\begin{equation}
R(M)_{ac}{}^{cb}+\df{2}{3}(*R(A))^b{}_{a}=0.
\end{equation}
This also follows from (\ref{eq:R-W}) which says that both
$R(M)_{ac}{}^{cb}$ and $R(A)_{ab}$ are given 
by $\na^\beta W_{\beta\alpha\gamma}$ and its conjugate.

The correspondence of the superconformal group transformations 
is as follows: 
\begin{equation}
\begin{array}{c|c}
\hline\hline
\text{component} & \text{superspace} 
\\ \hline
\delta_P(\xi^a)+\delta_{Q}(\varepsilon) & 
\delta_G(\xi(P)^a|P_a) + \delta_G(\xi(P)^{\ul{\alpha}}|Q_{\ul{\alpha}})
\,=\delta_G(\xi(P)^A|P_A)
\\ \hline
\delta_{{M}}(\lambda^{ab}) +\delta_{D}(\rho)+ \delta_{A}(\theta) & 
\delta_G(\Tspan{\tfrac12}\xi(M)^{ba}|M_{ab}) 
+ \delta_G(\xi(D)| D) + \delta_G(\xi(A)| A)
\\ \hline
\delta_{K}(\xi^a_K) + \delta_{S}(\zeta) &
\delta_G(\xi(K)^a| K_a) + \delta_G(\xi(K)^{\ul{\alpha}}| S_{\ul{\alpha}})
\,=\Tspan{\delta_G(\xi(K)^A| K_A)}
\\ \hline
\end{array}
\label{eq:SGT}
\end{equation}
The correspondence of transformation parameters is given in
(\ref{eq:parameter}). These can be shown by examining the commutation
relations in both approaches. The correspondence is trivial for 
the $M_{ab}$, $D$, $A$, $S$, $K_a$ transformations, but slightly
non-trivial for the commutation relations 
of $P_A= (P_a,\,Q_\alpha,\,\b{Q}^{\d\alpha})$. In particular, the
supercharge $Q_{\ul{\alpha}}$ is treated differently in both
approaches. In superspace approach, it is the spinor part of the
translation in superspace so that it is defined to be a combination of
the general coordinate and gauge transformations. In component
approach, the $Q$ transformation is defined to be the YM group law of
superconformal group though it is deformed by the curvature constraints.

Let us examine the commutation relations of the $P_A$ transformation
which is defined in Eq.~(\ref{eq:P_Atrf}) as
\begin{equation}
 \delta_G(\xi^AP_A)
=\delta_{\text{GC}}(\xi^M :=  \xi^A{E_A}^M)
-\delta_G(\xi^M{h_M}^{\cal{B}'}X_{\cal{B}'}). 
\end{equation}
We thus need the commutation relations between two GC transformations
in superspace and the GC and group transformations $X_{\cal{B}'}$
other than $P_A$. Noting that the field-independent pieces are the
flat indexed parameters $\xi^A$ and $\eta^A$, we find with a
straightforward calculation the following commutation relations:
{\allowdisplaybreaks%
\begin{align}
& [\delta_{\text{GC}}(\xi^B{E_B}^N),\,
\delta_{\text{GC}}(\eta^C{E_C}^L)] \nn
&\hspace{2em}
=\delta_{\text{GC}}(\xi^N\eta^L(\der_L{E_N}^A-\der_N{E_L}^A){E_A}^M), \nn
& [\delta_G(\xi^A{h_A}^{\cal{A}'}X_{\cal{A}'}),\,
\delta_{\text{GC}}(\eta^A{E_A}^M)] \nn
&\hspace{2em}
=\delta_G(\eta^L(\der_L\xi^N){h_N}^{\cal{B}'}X_{\cal{B}'})
+\delta_G(\eta^L\xi^N(\der_L {E_N}^A){h_A}^{\cal{B}'}X_{\cal{B}'}) \nn
&\hspace{4em}
-\delta_{\text{GC}}(\eta^L{E_L}^{C}\xi^N{h_N}^{\cal{B}'}
{f_{\cal{B}'C}}^D{E_D}^M), \\
& [\delta_G(\xi^A{h_A}^{\cal{A}'}X_{\cal{A}'}),\,
\delta_G(\eta^B{h_B}^{\cal{B}'}X_{\cal{B}'})] \nn
&\hspace{2em}
=\delta_G(\xi^L\eta^N({h_N}^{\cal{B}'}{E_L}^E-{h_L}^{\cal{B}'}{E_N}^E)
{f_{\cal{B}' E}}^F{h_F}^{\cal{A}'}X_{\cal{A}'}) \nn
&\hspace{4em}
+\delta_G((\eta^N(\der_N\xi^L)-\xi^N(\der_N\eta^L)){h_L}^{\cal{A}'}X_{\cal{A}'})
+\delta_G(\eta^N\xi^L {R_{LN}}^{\cal{A}'}X_{\cal{A}'}).  \nonumber
\end{align}}%
Using these relations and the definition of $P_A$ transformation, we obtain
\begin{equation}
[\delta_G(\xi^AP_A),\,\delta_G(\eta^BP_B)] = 
-\delta_G(\xi^A\eta^B{R_{AB}}^{\cal{C}}X_{\cal{C}}).
\label{eq:PP}
\end{equation}
The parameter $\xi^A$ is either vector $\xi^a$ or spinor
$\xi^{\ul{\alpha}}$. When we take both $\xi^A$ and $\eta^B$ to be
spinors, Eq.~(\ref{eq:PP}) implies the following $Q$-$Q$ commutation
relation by using the constraints on $R_{\ul{\alpha}\ul{\beta}}$,
\begin{equation}
[\delta_G(\xi^{\ul{\alpha}} Q_{\ul{\alpha}} ),\,
\delta_G(\eta^{\ul{\beta}}Q_{\ul{\beta}})]
=2\delta_G
\( \bmx \eta^\beta&\b\eta_{\d\beta} \emx
i\stog{}{a}{\beta}{\alpha}
\bmx \xi_\alpha \\
\b\xi^{\d\alpha} \emx
P_a \).
\end{equation}
This agrees with the $Q$-$Q$ commutation relation in component approach 
\begin{equation}
[\delta_Q(\varepsilon_1),\delta_Q(\varepsilon_2)]
=\delta_P\Big(\df{1}{2}\bar{\varepsilon}_2\gamma^a\varepsilon_1\Big),
\end{equation}
if $\tfrac12\b\varepsilon_1\corr\bmx\xi^\alpha&\b\xi_{\d\alpha}\emx|$
and $\tfrac12\b\varepsilon_2\corr\bmx\eta^\beta&\b\eta_{\d\beta}\emx|$
as given in the correspondence table (\ref{eq:parameter}). Next, if we
consider the vector parameter $\xi^a$ and the spinor parameter
$\eta^{\ul{\beta}}$, Eq.~(\ref{eq:PP}) becomes
\begin{equation}
\begin{split}
& [\delta_G(\xi^aP_a),\,\delta_G(\eta^{\ul{\beta}} Q_{\ul{\beta}})]  \\
& = -\(
\delta_G\bigl(\tfrac{1}{2}\xi^a\eta^{\ul{\beta}}{R(M)_{a{\ul{\beta}}}}
^{dc}M_{cd}\bigr)
+\delta_G\bigl(\xi^a\eta^{{\ul{\beta}}}{R(K)_{a{\ul{\beta}}}}^{{\ul{\gamma}}}
S_{\ul{\gamma}}\bigr)
+\delta_G\bigl(\xi^a\eta^{\ul{\beta}}{R(K)_{a{\ul{\beta}}}}^{c}K_{c}\bigr)\).
\end{split}
\end{equation}
Using the curvature expression $R_{a\beta}=
-i(\sigma_a)_{\beta\d\gamma}\cal{W}^{\d\gamma}$, this commutation
relation corresponds to
\begin{equation}
\begin{split}
[\delta_{P}(\xi^a),\,\delta_Q(\varepsilon)] & =
\sum_{A=M,S,K}\delta_A(\xi^b\delta'_Q(\varepsilon){h_b}^A)  \\
& =\delta_M\Big(\df{1}{2}\xi^c R^{ab}(Q)\gamma_c\varepsilon\Big) 
+\delta_S\Big(\df{1}{4}i\xi^a\gamma^b 
(\gamma_5 R_{ba}(A)+\tilde{R}_{ba}(A))\varepsilon\Big)  \\
& \qquad +\delta_K\Big(-\df{1}{2}\xi^bR^{\text{cov}}_{cb}(S)\sigma^{ac}
-\df{1}{4}\xi^b\delta^{ac}\tilde{R}^{\text{cov}}_{cb}(S)
\gamma_5\varepsilon\Big),
\end{split}
\end{equation}
in component approach 
when $\tfrac12\b\varepsilon=\bmx \eta^\beta & \b\eta_{\d\beta}\emx|$.
Finally, setting both $\xi^A$ and $\eta^B$ to be vectors, we have
\begin{equation}
[\delta_G(\xi^aP_a),\,\delta_G(\eta^bP_b)]
=-\delta_G(\xi^a\eta^b{R_{ab}}^{\cal{A}}X_{\cal{A}}), 
\end{equation}
which reproduces 
\begin{equation}
[\delta_{P}(\xi_1^a),\,\delta_{P}(\xi_2^b)]
=\sum_{A\neq P}\delta_A (\xi_1^a\xi_2^bR^{\text{cov}A}_{ab})
\end{equation}
in component approach with the correspondence $\xi^a_1\corr\xi^a|$ and
$\xi_2^b\corr\eta^b|$. Note that both $R_{ab}(P^c)$ in component approach
and ${R(P)_{ab}}^c$ in superspace vanish.

We remark the geometrical meaning of the correspondence of commutation
relations. In particular, the commutation relation 
$[\delta_{P},\delta_Q]$, which is algebraically determined by some 
constraints in component formulation, is understood as a vector-spinor
curvature in superspace. 

\subsection{Conformal multiplet}

We have shown that the superconformal transformations in both approaches 
satisfy exactly the same algebra. Once the algebra is fixed, the
transformation rule for a general conformal multiplet is uniquely
determined in component approach. That is, if the component with the
lowest Weyl weight is specified, all other components in the multiplet
and their transformation rules are found, up to some ambiguity in
field definitions. So we are lead to the exact correspondence of
superconformal multiplets
\begin{equation}
\hbox{Conformal multiplet}\ {\cal V}_\Gamma
\ \hbox{in (\ref{eq:cfmatter})} \ \ \corr \ \ 
\hbox{Primary superfield}\ \Phi_\Gamma \ \hbox{in (\ref{eq:Bpsf})}.
\nonumber
\end{equation}
In component approach, the first component $\cal{C}_\Gamma$ in
$\cal{V}_\Gamma$ is defined to have the lowest Weyl weight in the
multiplet so that its $S$ and $K_a$ transformations, which lower the
Weyl weight, must vanish. In superspace approach, a primary superfield
is defined to being $K_A$ invariant. As discussed before,
$\cal{C}_\Gamma$ and $\Phi_\Gamma|$ satisfy the same form of
superconformal transformations, Eq.~(\ref{eq:tlol}) and
Eq.~(\ref{eq:Bpsf}), respectively. Further if they have the same Weyl
weight $w=\Delta$ and chiral weight $n=(3/2)w$ as well as the same
representation matrices for Lorentz group $\Sigma_{ab}=-\cal{S}_{ab}$,
the multiplets in both approaches coincide with each other. The higher
components are determined successively by $Q$ transformations and some
ambiguity in field definitions are fixed by the standard form
(\ref{eq:sf}) in component approach~\cite{bib:KU}.

Thus in superspace approach, higher components in a superfield can be
found by applying $Q_{\ul{\alpha}}$ ($=\nabla_{\ul{\alpha}}$)
successively and comparing them with the transformation laws
in component approach. The detail is given in 
appendix \ref{sec:deriv_multi} from which we find the following
superfield expressions for the correspondence of a conformal multiplet
with the Weyl weight $w$ and the Lorentz index $\Gamma$
\begin{equation}
\begin{array}{c|c|l}
\hline\hline
\text{Weyl weight}& \text{component}&\multicolumn{1}{c}{\text{superspace}}\\
\hline
w & \cal{C}_\Gamma &
\ \Tspan{\Phi_\Gamma|}
\\ \hline
w+\tfrac12 & \cal{Z}_\Gamma &
\Tspan{
\bmx
-i\na_\alpha \Phi_\Gamma \\
+i{\b\na}^{\d\alpha}\Phi_\Gamma
\emx \!\Big| }  
\\ \hline
& \cal{H}_\Gamma &
\Tspan{ +\frac{1}{4}(\na^2 \Phi_\Gamma + \b\na^2 \Phi_\Gamma)|}
\\ \cline{2-3}
w+1 & \cal{K}_\Gamma & 
\Tspan{-\frac{i}{4}(\na^2 \Phi_\Gamma - \b\na^2 \Phi_\Gamma)|}
\\ \cline{2-3}
& \cal{B}_{a\Gamma} &
\Tspan{-\frac{1}{4}(\b\sigma_a)^{\d\beta\beta}
[\na_\beta,{\b\na}_{\d\beta}]\Phi_\Gamma| } 
\\ \hline
w+\tfrac32 & \Lambda_\Gamma & 
\Tspan{\df{i}{4}
\bmx
-\b\na^2\na_\alpha \Phi_\Gamma \\
+\na^2\b\na^{\d\alpha}\Phi_\Gamma
\emx \!\Big|
}
+2i
\bmx
\cal{W}_\alpha\\
{\cal{W}}^{\d\alpha}
\emx
\Phi_\Gamma|
\\
\hline
w+2& \cal{D}_\Gamma&
\Tspan{\tfrac{1}{8}\b\na_{\d\alpha}\na^2\b\na^{\d\alpha}\Phi_\Gamma|
+\cal{W}_{\d\alpha}\b\na^{\d\alpha}\Phi_\Gamma | } 
\\ & &
\qquad =
\Tspan{\tfrac{1}{8}\na^\alpha\b\na^2\na_\alpha \Phi_\Gamma|
-\cal{W}^\alpha \na_\alpha \Phi_\Gamma| }
\\
\hline
\end{array}
\label{eq:KUBcorr}
\end{equation}
In this correspondence, the overall factor is fixed by the
identification of the first components $\cal{C}_\Gamma\corr\Phi_\Gamma|$. 
In the last line, we have used an identity
\begin{equation}
\na^\alpha\b\na^2\na_\alpha
-\b\na_{\d\alpha}\na^2 \b\na^{\d\alpha}
= 8\big( \cal{W}_{\d\alpha}\b\na^{\d\alpha}
+ \cal{W}^\alpha \na_\alpha 
+\{\b\na_{\d\alpha},\,\cal{W}^{\d\alpha}\} \big),
\label{eq:PTid}
\end{equation}
which is the conformal superspace counterpart of the identity 
$D^\alpha \b{D}^2D_\alpha-\b{D}_{\d\alpha}D^2\b{D}^{\d\alpha} = 0$
in global supersymmetry. The RHS in (\ref{eq:PTid}) comes from nonzero
vector-spinor curvatures and depends on the gaugino superfield
$\cal{W}_{\ul{\alpha}}$. Noticing that $\cal{W}_\alpha$ has only the
$M$ and $K_A$ components (see Eq.~(\ref{eq:gaugino})), the above
superfield expressions of $\Lambda$ and $\cal{D}$ for a multiplet with
no Lorentz index reduce to
\begin{equation}
\Lambda \ \ \corr\ \ \df{i}{4}
\bmx
-\b\na^2\na_\alpha \Phi \\
+\na^2\b\na^{\d\alpha}\Phi
\emx\!\Big|,
\qquad\quad
\cal{D} \ \ \corr\ \
\df{1}{8}{\b\nabla}_{\d\alpha} \na^2 \b\na^{\d\alpha}\Phi|
= \df{1}{8} \na^\alpha\b\na^2 \na_\alpha \Phi|,
\end{equation}
which are the same forms as in global supersymmetry.

\subsection{Chiral projection and invariant actions}
\label{sec:proj-act}

In this subsection we discuss the correspondences of chiral multiplets,
the chiral projection and the superconformally invariant actions.

In superspace approach, a primary chiral superfield $\Phi_\Gamma$ is
defined to be a primary superfield satisfying the chirality condition
\begin{equation}
\b\na_{\d\alpha} \Phi_\Gamma=0 \,.
\end{equation}
Since ``primary'' means the $K_A$ invariant, a consistency for such
multiplet to exist requires
\begin{equation}
0 = \{\b{S}^{\d\alpha},\b\na^{\d\beta}\}\Phi_\Gamma
=\((2D+3iA)\epsilon^{\d\alpha\d\beta}-2M^{\d\alpha\d\beta}\)\Phi_\Gamma,
\end{equation}
which demands $\Phi_\Gamma$ to have the Weyl and chiral weights 
$(\Delta, w)$ satisfying $2\Delta-3w=0$ and to carry only undotted
spinor indices $\Gamma=(\alpha_1\alpha_2\cdots)$. These conditions for
weights and Lorentz index are exactly the same as given in
Eq.~(\ref{eq:chiral_comp}) in component approach. The component fields
in a conformal multiplet with the chirality condition are found from
the correspondence table (\ref{eq:KUBcorr})
\begin{equation}
\bigl( \cal{C}_\Gamma, \cal{Z}_\Gamma, \cal{H}_\Gamma, 
\cal{K}_\Gamma, \cal{B}_{a\Gamma}, \Lambda_\Gamma, \cal{D}_\Gamma \bigr)
\ \ \corr \ \
\(\Phi_\Gamma|,\, \na_\alpha\Phi_\Gamma|,\, \tfrac14\na^2\Phi_\Gamma|, 
\,\tfrac{-i}{4}\na^2\Phi_\Gamma|,\, i\na_a\Phi_\Gamma|,\, 0,\, 0\) ,
\label{eq:embedding}
\end{equation} 
by using the equations 
\begin{equation}
\b\na_{\d{\beta}}\na_\alpha\Phi_\Gamma=
\{\b\na_{\d{\beta}},\,\na_\alpha\}\Phi =-2i\na_{\alpha\d\beta}\Phi_\Gamma, 
\qquad \cal{W}^{\d\alpha}\Phi_\Gamma=0. 
\end{equation}
The last equation follows from $M^{\d\beta\d\gamma}\Phi_\Gamma=
K_A\Phi_\Gamma=0$ for a primary superfield $\Phi_\Gamma$ with purely
undotted $\Gamma$. Comparing the expression (\ref{eq:embedding}) with
the embedding formula referred to above Eq.~(\ref{eq:222}) in
component approach, we find the following correspondence between a
conformal chiral multiplet in component approach and a primary chiral
superfield $\Phi_\Gamma$,
\begin{equation}
[\,\cal{A}_\Gamma,\ \cal{P}_{\text{R}}\chi_\Gamma,\ \cal{F}_\Gamma 
\,] \ \ \hbox{(component)} 
\ \ \corr\ \
\bigl[\,\Phi_\Gamma|,\ \na_\alpha\Phi_\Gamma|,\ -\tfrac14 \na^2\Phi_\Gamma| 
\;\bigr] \ \
\hbox{(superspace)} 
\label{eq:chiral-corr}
\end{equation}

The algebra $\{\b\na_{\d\alpha},\,\b\na_{\d\beta}\}=0$ 
in Eq.~(\ref{eq:nablaAlgebra}) implies the equation
\begin{equation}
\b\na_{\d\alpha} \b\na^2 \Psi_\Gamma=0 
\end{equation}
identically holds for any superfield $\Psi_\Gamma$. So 
$\b\na^2\Psi_\Gamma$ formally seems a chiral superfield. However,
if $\b\na^2\Psi_\Gamma$ is not primary, it still has to contain
$8+8$ components contrary to the fact that a chiral superfield has
only $2+2$ components. This odd property happens in the superconformal
case since $\b{S}_{\d\alpha}$ acts as an inverse operator of
$\b\na_{\d\alpha}$.

If $\b\na^2\Psi_\Gamma$ is primary, it contains only $2+2$ components
for a primary chiral superfield. For $\Psi_\Gamma$ with the Weyl and
chiral weights $(\Delta,w)$, $\b\na^2\Psi_\Gamma$ has 
$(\Delta+1,w+2)$ and becomes chiral and primary if
$2(\Delta+1)-3(w+2)=0$ and $\Gamma$ is purely undotted. This means
that $\b\na^2$ gives a chiral projection operator if it acts on a
primary superfield $\Psi_\Gamma$ whose weights and index satisfy those
conditions. That agrees with the conditions for the chiral projection 
operator in component approach given in
Eq.~(\ref{eq:chiralproj}). Taking care of coefficients, we find the
correspondence between the chiral projection operators $\Pi$ in
component approach and $\cal{P}$ in superspace 
\begin{equation}
\Pi  \ \ \corr \ \ -\cal{P}=\df{1}{4}\b\na^2.
\label{eq:proj-corr}
\end{equation}
We show in appendix \ref{sec:deriv_proj} that the component fields of
a projected superfield $\cal{P}\Psi_\Gamma$ are identified with those
of $\Pi\cal{V}_\Gamma$ in Eq.~(\ref{eq:chiralproj}) in component
approach. In this identification, the following equations are useful
\begin{equation}
\begin{split}
\na^2\b\na^2
&=\b\na_{\d\alpha}\na^2\b\na^{\d\alpha}+8\na^a\na_a
-2i\na_a(\b\sigma^a)^{\d\alpha\alpha}[\na_\alpha,\b\na_{\d\alpha}]
-8\cal{W}_{\d\alpha}\b\na^{\d\alpha}, \\[1mm]
\b\na^2\na^2
&=\na^\alpha\b\na^2\na_\alpha +8\na^a\na_a
+2i\na_a(\b\sigma^a)^{\d\alpha\alpha}[\na_\alpha,\b\na_{\d\alpha}]
+8\cal{W}^\alpha \na_\alpha.
\end{split}
\label{eq:ChiralIdentity}
\end{equation}
We remark that the sum of these yields
\begin{equation}
\na^2 \b\na^2 +\b\na^2\na^2 -\na^\alpha \b\na^2 \na_\alpha
-\b\na_{\d\alpha}\na^2 \b\na^{\d\alpha}
=16\na^a\na_a +8\cal{W}^\alpha\na_\alpha
-8\cal{W}_{\d\alpha}\b\na^{\d\alpha}, 
\end{equation}
which is the conformal superspace counterpart of the global
supersymmetry identity
\begin{equation}
D^2\b{D}^2+\b{D}^2D^2-2D^\alpha \b{D}^2D_\alpha =16\square\,.
\end{equation}

Finally we discuss the superconformally invariant actions. First is
the correspondence of the F-type invariant action for the conformal
chiral multiplet $\Sigma$ without external Lorentz index. The
component expansion of the F-type integration (\ref{eq:F-int}) is
coincident with the expression (\ref{eq:F-action}) of F-type invariant
action in component approach, if taken account of the correspondences
of gauge fields (\ref{eq:GF}) and chiral multiplet components
(\ref{eq:chiral-corr})
\begin{equation}
\int d^4 x \bigl[\,\Sigma\,\bigr]_F \ \ \corr\ \  
\int d^4 x d^2 \theta \, \cal{E}\Sigma +
\int d^4 x d^2 \b\theta \,\b{\cal{E}}\b\Sigma.
\label{eq:F-corr}
\end{equation}
The other is the correspondence of the D-type invariant action for the
general real conformal multiplet $V$ without external Lorentz
index. Since the D-type formula is obtained from the F-type one by
using the chiral projection operator, the correspondences of
(\ref{eq:proj-corr}) and (\ref{eq:F-corr}) directly lead to the
following correspondence of the D-type invariant actions
(\ref{eq:D-action}) in component approach and the D-type integral
(\ref{eq:DtoF}) in superspace approach
\begin{equation}
\int d^4 x \bigl[\,V\,\bigr]_D \ \ \corr\ \ 2\int d^4 x d^4\theta\, EV.
\label{eq:DD-corr}
\end{equation}

\subsection{$\bs{u}$-associated derivatives}

\subsubsection{restriction for the existence of conformal spinor derivatives?}

We first mention to a historical puzzle on the conformal spinor
derivative. In Ref.~\cite{bib:KU}, KU constructed the spinor
derivative in component approach and claimed that such spinor
derivative $\scr{D}_{\ul{\alpha}}$ exists only when some special
conditions are met on an operand multiplet $\cal{V}_\Gamma$. On the
other hand, Butter defined~\cite{bib:B1} in superspace formalism the
conformally covariant derivatives $\na_A$ which can act on any
superfield with no restriction. What is the difference?

The point is that KU defined in their component approach a conformal
multiplet $\cal{V}_\Gamma$ by its first component $\cal{C}_\Gamma$,
denoting $\cal{V}_\Gamma=\cfbra \cal{C}_\Gamma \cfket$, which has the
lowest Weyl weight in the multiplet. Therefore the $S$ and $K_a$
transformations of $\cal{C}_\Gamma$ must vanish since $S$ and $K_a$
lower the Weyl weight. In superspace terminology, such a multiplet is
arranged in a primary superfield $\Phi_\Gamma$:
\begin{equation}
\cal{V}_\Gamma=\cfbra \cal{C}_\Gamma \cfket \ \ \corr\ \ \Phi_\Gamma,  
\qquad\quad 
\delta_S\cal{C}_\Gamma=\delta_K\cal{C}_\Gamma=0 \ \ \corr\ \ K_A\Phi_\Gamma=0.
\end{equation} 
KU looked for the spinor derivative $\scr{D}_{\ul{\alpha}}$ as a
mapping of a conformal multiplet $\cal{V}_\Gamma$ to another conformal
multiplet whose first component is $\cal{Z}_{\ul{\alpha}\Gamma}$ which
is the second component of $\cal{V}_\Gamma$
\begin{equation}
\scr{D}_\alpha \;:\; 
\cal{V}_\Gamma=\cfbra \cal{C}_\Gamma \cfket \ \ \rightarrow\ \ 
\scr{D}_\alpha\cal{V}_\Gamma=\cfbra \cal{Z}_{\alpha\Gamma} \cfket\,.
\end{equation}
That is crucial and only difference from the superspace covariant
derivatives $\nabla_A$, which generally do not bring a primary
superfield into primary. This freedom employed in superspace
formulation is consistent with the freedom of $Q$ transformation in
component approach, because the $S$ transformation of
$\cal{Z}_{\alpha\Gamma}$ is not generally required to vanish. Thus the
conformal covariant spinor derivative that corresponds to the $Q$
transformation is $\nabla_{\ul{\alpha}}$, not
$\scr{D}_{\ul{\alpha}}$. Conversely speaking, once the image
$\na_\alpha\Phi_\Gamma$ is required to be 
primary, $S_\beta \na_\alpha\Phi_\Gamma =0$ leads to the same
conditions for $\Phi_\Gamma$ as KU found in component approach.

\subsubsection{$\bs{u}$-associated derivative}

We need the $S$ and $K_a$ invariance of multiplets, for instance, in
constructing the invariant actions by the D-type and F-type
formulas. Ref.~\cite{bib:KU} has shown that, if one has a compensating
multiplet $\bs{u}$ (or any conformal multiplet whose first component
is guaranteed to be non-vanishing, like the compensator used for gauge
fixing), the covariant derivative ${\cal{D}_\alpha}^{(\bs{u})}$ is
constructed which maps a conformal multiplet into another conformal
one without any restriction.

Consider a conformal multiplet $\bs{u}$ with the Weyl and chiral weights 
$(w_0,n_0)$ and no external Lorentz index. The component fields are
denoted as
\begin{equation}
\bs{u} = \left[\,\cal{C}_u,\cal{Z}_u,\cal{H}_u,\cal{K}_u,
{\cal{B}_u}_a,\Lambda_u,\cal{D}_u\,\right].
\end{equation}
Assuming that the first component $\cal{C}_u$ is non-vanishing, we
construct the following spinor
\begin{equation}
\lambda^S:=\df{i\cal{Z}_u}{(w_0+n_0)\cal{C}_u},
\end{equation}
which is non-linearly shifted under the $S$ transformation as 
$\delta_S(\zeta)\lambda^S=\zeta$. Then the $\bs{u}$-associated spinor
derivative ${\cal{D}_\alpha}^{(\bs{u})}$ is defined by
\begin{equation}
{\cal{D}_\alpha}^{(\bs{u})}\cal{V}_\Gamma =
\Bigbra \cal{Z}_{\alpha\Gamma}+i(w+n){\lambda_\alpha}^S\cal{C}_\Gamma
-{(\sigma_{ab})_\alpha}^{\beta} {\lambda_\beta}^S(\Sigma^{ab}\cal{C})_\Gamma
\Bigket \,,
\label{eq:u-assoc-spinder}
\end{equation}
where $w$ and $n$ are the Weyl and chiral weights of
$\cal{C}_\Gamma$. Since $\delta_S(\zeta) \cal{Z}_\Gamma=
-i(w+n)\zeta\cal{C}_\Gamma +\sigma_{ab}\zeta\,(\Sigma^{ab}\cal{C})_\Gamma$, 
the quantity in the bracket on the RHS is invariant under the $S$
transformation, so that it defines a conformal multiplet 
${\cal{D}_\alpha}^{(\bs{u})}\cal{V}_\Gamma$. The barred derivative
$\b{\cal{D}}^{\d\alpha(\bs{u})}$ is given by 
$\b{\cal{D}}^{\d\alpha (\bs{u})}\cal{V}_\Gamma
=(\cal{D}^{\alpha(\bs{u})}\cal{V}_\Gamma^* )^*$.

Similarly, the $\bs{u}$-associated vector derivative is constructed as
follows. We define a vector $V_a{}^K$ and a spinor $\chi^S$ by 
\begin{equation}
{V_a}^K := \df{1}{4w_0}\(\df{D_a \cal{C}_u}{\cal{C}_u}
+\df{D_a{\cal{C}_u}^*}{{\cal{C}_u}^*}\),  \qquad
\chi^S:=\df{1}{2w_0}i\gamma_5 \(\df{\cal{Z}_u}{\cal{C}_u}
+\df{{\cal{Z}_u}^*}{{\cal{C}_u}^*}\) \,,
\end{equation}
so that ${V_a}^K$ and $\chi^S$ are shifted under the $K_a$ and $S$
transformations, respectively, as $\delta_K(\xi^K){V_a}^K= \xi_a^K$ and 
$\delta_S(\zeta) \chi^S =\zeta$. The $S$ transformation of the vector
${V_a}^K$ yields the spinor $\chi^S$ as 
$\delta_S(\zeta) {V_a}^K=\tfrac{-1}{4}\bar{\zeta}\gamma_a \chi^S$. By
adding appropriate terms containing these fields, the superconformally
covariant derivative $D_a\cal{C}_\Gamma$ defined in
(\ref{eq:cfcovder}) can be made $S$-invariant, and the
$\bs{u}$-associated vector derivative is defined by
\begin{equation}
\begin{split}
{\cal{D}_a}^{(\bs{u})} \cal{V}_\Gamma & =\Bigbra 
D_a\cal{C}_\Gamma-2w{V_a}^K\cal{C}_\Gamma+
2V^{bK}(\Sigma_{ab}\cal{C})_\Gamma
+\df{1}{2}\b\chi^S\gamma_a i\gamma_5\cal{Z}_\Gamma  \\
& \hspace{8em} +\df{1}{4}(\b\chi^S\gamma_5
\gamma^b\chi^S)\bigl(\delta_{ab}n\cal{C}_\Gamma
+(\tilde{\Sigma}_{ab}\cal{C})_\Gamma\bigr) \Bigket \ ,
\end{split}
\label{eq:u-assoc-vecder}
\end{equation}
so that ${\cal{D}_a}^{(\bs{u})}\cal{V}_\Gamma$ is a conformal multiplet.

We now show the superspace expression for the $\bs{u}$-associated
derivatives using the correspondences given in the previous
section. First we introduce the primary superfield $X_u$ that
corresponds to $\bs{u}$
\begin{equation}
\bs{u}\ \ \corr\ \  X_u,
\end{equation}
where $X_u$ has the Weyl and chiral weights $(\Delta_0,w_0)=
(w_0,\frac{2}{3}n_0)$. From the correspondences of weights and
component fields (\ref{eq:KUBcorr}), $\lambda^S$ is identified as
\begin{equation}
{\lambda_\alpha}^S \ \ \corr\ \ 
\df{2}{(2\Delta_0+3w_0)X_u|}\nabla_\alpha X_u|.
\end{equation}
By reading the correspondence of ${\cal{D}_\alpha}^{(\bs{u})}\cal{V}_\Gamma$ 
(\ref{eq:u-assoc-spinder}), we find the following 
superspace expression for the $u$-associated spinor derivative
\begin{equation}
{\cal{D}_\alpha}^{(\bs{u})}\cal{V}_\Gamma \ \ \corr\ \
-i\Bigl(\na_\alpha + \df{1}{(2\Delta_0+3w_0)X_u}(\na^\beta X_u)
\{S_\beta,Q_\alpha\}\Bigr)\Phi_\Gamma,
\label{eq:Dau_sp}
\end{equation}
and similarly for the dotted spinor derivative. When $X_u$ is a real
superfield with special weights $\Delta_0=2$ and $w_0=0$, this
expression reduces to the compensated spinor derivatives discussed in
Ref.~\cite{bib:B2}. So, (\ref{eq:Dau_sp}) stands for the
generalization to $X_u$ with arbitrary weights.

We also construct the superspace expression for the
$\bs{u}$-associated vector derivative. For this purpose, we consider
the real superfield $Y_u$ defined by\footnote{Precisely speaking, this
$Y_u$ itself is not a proper primary superfield unless
$\Delta_0=w_0=0$ since $\log X_u$ has no definite values of Weyl and
chiral weights. In the following expressions, however, only its
derivative $\na_A Y_u = {\na_A X_u}/{X_u}+{\na_A \b{X}_u}/{\b{X}_u}$ 
appears, which is a proper superfield with the Weyl and chiral weights 
of the operator $\na_A$.}
\begin{equation}
Y_u = \log X_u+\log \b{X}_u.
\end{equation}
Using the component field correspondence (\ref{eq:KUBcorr}), we identify 
${V_a}^K$ and $\chi^S$ as
\begin{equation}
{V_a}^K \ \corr\ \ \df{1}{4\Delta_0}\na_a Y_u|,
\qquad\quad
\chi^S \ \corr\ \ {\df{1}{2\Delta_0}}i\axg 
\bmx
-i\na_\alpha Y_u \\
+i\b\na^{\d\alpha} Y_u
\emx| \ .
\end{equation}
Then the superspace expression for the $\bs{u}$-associated vector
derivative is found by translating (\ref{eq:u-assoc-vecder}):
\begin{eqnarray}
\begin{split}
{\cal{D}_a}^{(\bs{u})}\cal{V}_\Gamma \ \ \corr\ \ &
\na_a \Phi_\Gamma -\df{1}{2\Delta_0}(\na_a Y_u)D\Phi_\Gamma
-\df{1}{2\Delta_0} (\na^b Y_u)M_{ab}\Phi_\Gamma \\
& +\df{1}{4\Delta_0}
\bmx 
\na^\alpha Y_u & \b\na_{\d\alpha}Y_u
\emx
i\stog{a}{}{\alpha}{\beta} i\axg
\bmx
-i\na_\beta \Phi_\Gamma \\
+i\b\na^{\d\beta}\Phi_\Gamma
\emx  \\
& \quad\qquad
+\df{1}{8\Delta_0}
\bmx
\na^\alpha Y_u & \b\na_{\d\alpha} Y_u
\emx
\axg i\stog{}{b}{\alpha}{\beta} \df{1}{2\Delta_0}
\bmx
\na_\beta Y_u \\
\b\na^{\d\beta}Y_u
\emx\\
& \quad\qquad
\times \(-\df{3i}{2}\eta_{ab}A
+\df{1}{2}(-i\varepsilon_{abcd})(-M^{cd})\)\Phi_\Gamma.
\end{split}
\end{eqnarray} 
When $X_u$ is a real primary superfield $X$ with the weights
$(\Delta_0,w_0)=(2,0)$, this reduces to the compensated vector
derivative with parameter $\lambda=1$ given in Ref.~\cite{bib:B2}, if
we replace $Y_u \rightarrow 2\log X$.

\section{Superconformal gauge fixing to Poincar\'e SUGRA}
\label{sec:scym}

The superconformal group is larger than the super Poincar\'e and the
extra $D$, $A$, $S$, $K_a$ gauge symmetry should be fixed to have
Poincar\'e SUGRA, which is useful e.g.\ for phenomenological
applications. In this section, we examine the superconformal gauge
fixing of the matter-coupled conformal SUGRA to Poincar\'e SUGRA, and
give the correspondence of gauge-fixing conditions between superspace
and component approaches. In this paper we focus on the chiral
superfield matter system. The gauge fixing for the system containing
YM gauge fields of internal symmetry will be discussed
elsewhere~\cite{bib:KYY2}.

\subsection{Gauge fixing in superspace approach}

The matter superfields $\Phi^i$ ($i=1,2,\ldots,n$) are introduced to
be primary and covariantly chiral with respect to the superconformal 
symmetry, $\b\na^{\d\alpha}\Phi^i=0$. They have the Weyl and chiral
weights $(\Delta,w)=(0,0)$. The matter-coupled SUGRA action in
conformal superspace is given by
\begin{equation}
S = -3\int d^4x d^4\theta\, E\,\Phi^\co\b\Phi^\co e^{-K/3}
+\(\int d^4x d^2 \theta\, \cal{E}\,(\Phi^\co)^3W +\text{h.c.}\),
\end{equation}
where the K\"ahler potential $K=K(\Phi^i,\b\Phi^{i^*})$ is a real
function of matter superfields, and the superpotential $W=W(\Phi^i)$
is a holomorphic one. In the first term (the D-type action), the
superconformal gauge invariance leads to the conditions that the
compensator chiral superfield $\Phi^\co$ is primary and has the
weights $(\Delta,w)=(1,2/3)$. In the second term (the F-type action),
the compensator dependence is also fixed by the superconformal gauge
invariance.

Let us discuss the gauge fixing of superconformal symmetry to go down
to Poincar\'e SUGRA\@. For a non-vanishing superpotential, it is
useful to redefine the compensator $\Phi^\co$ as\footnote{The
redefinition (\ref{eq:redef}) is possible when $W\neq0$. For $W=0$, a
convenient gauge choice may be $\Phi^\co=e^{K/6}$ (and $B_M=0$) which
is the same condition as the one given in \cite{bib:B1} and mentioned
in section \ref{sec:conf_ss}.}
\begin{equation}
\Phi^\co \ \rightarrow\ \Phi^0 = \Phi^\co W^{1/3}.
\label{eq:redef}
\end{equation}
The new chiral compensator $\Phi^0$ has the weights 
$(\Delta,w)=(1,2/3)$. The action in terms of $\Phi^0$ is given by
\begin{equation}
S = -3\int d^4x d^4\theta\, E\, \Phi^0\b\Phi^0 e^{-G/3}
+\(\int d^4xd^2\theta\, \cal{E}\, (\Phi^0)^3+\text{h.c.}\)
\end{equation} 
with 
\begin{equation}
G = K + \ln |W|^2.
\end{equation}
One of the virtues of using $\Phi^0$ and $G$ is revealed in
introducing YM gauge fields, that is, $\Phi^0$ and $G$ are invariant
under possible internal symmetry, while $\Phi^\co$ and $K$ are not
invariant. This invariant property of $\Phi^0$ and $G$ makes it simple
to fix the superconformal gauge symmetry irrespectively of internal
ones~\cite{bib:KYY2}.

In component approach, Ref.~\cite{bib:KUigc} discussed the
superconformal gauge-fixing conditions which realize the canonically
normalized EH and RS terms and also give a real gravitino mass, given
in (\ref{eq:KUgaugeCond}). We find its superspace counterparts are 
\begin{equation}
D,\ A\hbox{-gauge}:\ \Phi^0=\bar{\Phi}^0=e^{G/6}, \qquad
K_A\hbox{-gauge}:\ B_M=0.
\label{eq:GF-G}
\end{equation}
The second condition is imposed by an appropriate $K_A$ gauge
transformation of the $D$-gauge superfield: 
$\delta_G(\xi(K)^AK_A)B_M=-2{E_{MA}}\xi(K)^A$. On the other hand, the 
first condition seems peculiar since the chiral superfield $\Phi^0$ 
does not have enough numbers of independent components which can be
set equal to the general real superfield $e^{G/6}$. It is however
noticed that the gauge fixing (\ref{eq:GF-G}) is given in conformal
superspace where all gauge transformations have real superfield
parameters. Therefore the finite $D$ and $A$ gauge transformations
$\Phi^0\mapsto e^{\xi(D)+\frac{2i}{3}\xi(A)}\Phi^{0}$ are possible
with the real superfield parameters 
$\xi(D)={G/6}-(1/2)\ln(\Phi^0\b\Phi^0)$ and 
$\xi(A)=(3/4i)\ln(\b{\Phi}^0/\Phi^0)$ which brings $\Phi^0$ to $e^{G/6}$. 

The gauge-fixing conditions (\ref{eq:GF-G}) imply several other
equations for superfield components. We here focus on the chiral
compensator $\Phi^0$ and the $A$-gauge superfield $A_M$. Recall that
the covariant derivative takes the following form for a primary
superfield $\Phi^{(\Delta,w)}$ with the Weyl and chiral weights
$(\Delta,w)$ and no external Lorentz index:
\begin{align}
\na_A\Phi^{(\Delta,w)} &= \cal{D}^{\rm P}_A \Phi^{(\Delta,w)}
-\bigl(\Delta B_A+iw A_A\bigr) \Phi^{(\Delta,w)},
\label{eq:Nformula}  \\
\na_A\na_\alpha\Phi^{(\Delta,w)} &= 
\cal{D}^{\rm P}_A \na_\alpha\Phi^{(\Delta,w)} 
-\Bigl( \big(\Delta+\df12\big) B_A +i(w-1)A_A \Bigr)
\na_\alpha\Phi^{(\Delta,w)}  \nonumber \\
&\hspace{8em}
+(2\Delta+3w)f_{A\alpha}\Phi^{(\Delta,w)}, 
\label{eq:N2formula}
\end{align}
where the last term $(2\Delta+3w)f_{A\alpha}\Phi^{(\Delta,w)}$ stands
for $-f_A{}^\beta\{ K_\beta,\,\na_\alpha\}\Phi^{(\Delta,w)}$ and we
have used the equation $K_B\na_\alpha\Phi^{(\Delta,w)}
=[K_B, \na_\alpha]\Phi^{(\Delta,w)}=0$ for $B=b,\d\beta$ which is
satisfied if $\Phi^{(\Delta,w)}$ is primary. The derivative 
$\cal{D}^{\rm P}_A$ is defined by
\begin{equation}
\cal{D}^{\rm P}_A =
E_A{}^M\partial_M  -\df{1}{2}\phi_A{}^{bc}M_{cb} ,
\label{eq:PoincareDer}
\end{equation}
which is the covariant derivative in Poincar\'e SUGRA and different
from the derivative in Ref.~\cite{bib:B1} ($\cal{D}_A$ discussed in
section~\ref{sec:conf_ss}). Plugging the gauge-fixing conditions
(\ref{eq:GF-G}) into the RHS of (\ref{eq:Nformula}) and
(\ref{eq:N2formula}), we find the components of the chiral compensator
superfield $\Phi^0$,
\begin{align}
\Phi^0\big| &= e^{G/6}\big|,
\label{eq:phi0}  \\
\na_{\alpha}\Phi^0 \big| &= \Bigl(\cal{D}^{\rm P}_{\,\alpha} 
-\df{2}{3}iA_{\alpha}\Bigr) e^{G/6}\Bigr|,
\label{eq:aphi0}  \\
\na^2\Phi^0\big| &= \Bigl(\cal{D}^{\rm P}{}^\alpha+\df13iA^{\alpha}\Bigr) 
\Bigl(\cal{D}^{\rm P}_{\,\alpha}-\df{2}{3}iA_{\alpha}\Bigr) e^{G/6}\Big|
-4f_\alpha{}^\alpha e^{G/6}\Big|.
\label{eq:aaphi0}
\end{align}
Note that $\na_\alpha\Phi^0\not= \na_\alpha e^{G/6}$ but
$\cal{D}^{\rm P}_{\,\alpha}\Phi^0= \cal{D}^{\rm P}_{\,\alpha} e^{G/6}$
since the gauge-fixing condition $\Phi^0=e^{G/6}$ violates the $D$ and $A$
symmetries but preserves $M_{ab}$.

After the gauge fixing, the chirality condition of the compensator
$\Phi^0$ turns out to determine the $A$-gauge superfield. Applying
(\ref{eq:Nformula}) to $\b\na_{\d\alpha}\Phi^0=0$ and using the
gauge-fixing condition (\ref{eq:GF-G}), we obtain
\begin{equation}
0 = \b\na_{\d\alpha}\Phi^0=\b{\cal{D}}^{\rm P}_{\,\d\alpha}\Phi^0
-\df{2}{3}iA_{\d\alpha} \Phi^0
\quad \rightarrow \quad 
A_{\d\alpha} = 
-\df{i}{4}G_{j^*}\b{\cal{D}}^{\rm P}_{\,\d\alpha}\b{\Phi}^{j^*}, 
\end{equation}
where the field derivatives of $G$ are denoted as
$G_i=\partial G/\partial\Phi^i$ and 
$G_{i^*}=\partial G/\partial\b{\Phi}^{i^*}$.
Similarly, the condition $\na_\alpha\b{\Phi}^0=0$ fixes $A_\alpha$ as
\begin{equation}
A_{\alpha} = \df{i}{4}G_j \cal{D}^{\rm P}_{\,\alpha}{\Phi}^{j},
\label{eq:Aalpha}
\end{equation}
with which the components of the chiral compensator, (\ref{eq:aphi0})
and (\ref{eq:aaphi0}), are rewritten as
\begin{align}
\na_\alpha\Phi^0\big| &=
\df{1}{3}e^{G/6}G_i \cal{D}^{\rm P}_{\,\alpha}\Phi^i\big|,
\label{eq:342}  \\
\na^2\Phi^0\big| & =\df13 e^{G/6} \Bigl(
G_i\cal{D}^{\rm P}{}^2\Phi^i +\bigl(G_{ij}+\df{1}{12}G_iG_j\bigr)
(\cal{D}^{\rm P}{}^\alpha\Phi^j)(\cal{D}^{\rm P}_{\,\alpha}\Phi^i) 
-24\b{R} \Bigr)\Bigr|.
\label{eq:341}
\end{align}
We have used the relation $f_{\alpha\beta}=
-\epsilon_{\alpha\beta}\b{R}$ which comes from the curvature
constraints after the gauge fixing. The equation (\ref{eq:341})
relates the compensator $F$ component to the auxiliary field $\b{R}$,
undetermined part of the $S$-gauge field $f_\alpha{}^\beta$. It is
noticed that superfield components are not given by the covariant
derivative of Poincar\'e SUGRA ($\cal{D}^{\rm P}{}$) but should be
defined by the conformal one ($\na$). For the matter superfields
$\Phi^i$ with vanishing weights $(\Delta,w)=(0,0)$, these two
derivatives give same results for the first derivatives (spinor
components), but different for the second ones ($F$ components). For
the comparison with component approach, we rewrite the above results
with the conformally covariant derivative $\nabla$. 
Eqs.~(\ref{eq:Nformula}), (\ref{eq:N2formula}) and (\ref{eq:Aalpha})
imply $\cal{D}^{\rm P}_{\,\alpha}\Phi^i=\na_\alpha\Phi^i$ and
\begin{align}
\cal{D}^{\rm P}{}^2\Phi^i
= \na^2\Phi^i -i A^\alpha\cal{D}^{\rm P}_{\,\alpha}\Phi^i
= \na^2\Phi^i 
+\df{1}{4}G_j \nabla^\alpha\Phi^j\nabla_\alpha\Phi^i.
\end{align}
We then find (\ref{eq:342}) and (\ref{eq:341}) are given by
\begin{align}
\na_\alpha\Phi^0\big| &=
\df{1}{3}e^{G/6}G_i \nabla_\alpha\Phi^i\big|, 
\label{eq:fix_spinor} \\
\na^2\Phi^0\bigr| &= \df13 e^{G/6} \Bigl(
G_i\na^2\Phi^i +\bigl(G_{ij}+\df{1}{3}G_iG_j\bigr)
\nabla^\alpha\Phi^j \nabla_\alpha\Phi^i -24\b{R} \Bigr)\Bigr|.
\label{eq:hcompensator}
\end{align}
The chirality condition of the compensator also fixes the vector part
of $A$-gauge field. The chirality condition and the algebra 
$\{\na_\alpha, \b{\na}^{\d{\beta}}\}=-2i\na_\alpha{}^{\d{\beta}}$ imply
\begin{equation}
\b\na^{\d\beta}\na_\alpha\Phi^0 = -2i\na_\alpha{}^{\d\beta}\Phi^0 .
\label{eq:CovChiral}
\end{equation}
After the gauge fixing, the vector derivative on the RHS becomes
$\na_\alpha{}^{\d\beta}\Phi^0=
\cal{D}^{\rm P}_{\,\alpha}{}^{\d\beta}\Phi^0 
-\frac{2}{3}iA_\alpha{}^{\d\beta}\Phi^0$ which is used to determine
the vector part $A_\alpha{}^{\d\beta}$. Evaluating the LHS of
(\ref{eq:CovChiral}) by using (\ref{eq:N2formula}) with the
gauge-fixing conditions (\ref{eq:GF-G}) and Eq.~(\ref{eq:342}), we find
\begin{equation}
\begin{split}
A_\alpha{}^{\d\beta} &= 
-\df{i}4 \cal{D}^{\rm P}_{\,\alpha}{}^{\d\beta} G 
-\df14 e^{-G/6}\Bigl(\b{\cal{D}}^{\rm P}{}^{\d\beta}+\df13iA^{\d\beta}\Bigr) 
e^{G/6}G_i \cal{D}^{\rm P}_{\,\alpha}\Phi^i
-3f^{\d\beta}{}_\alpha  \\ 
&= \df{i}{4}(G_i \na_\alpha{}^{\d\beta} \Phi^i
-G_{i^*}\na_\alpha{}^{\d\beta}
\b\Phi^{i^*})+\df{1}{4}G_{ij^*}\na_\alpha\Phi^i
\na^{\d\beta}\b\Phi^{j^*}-\df{3}{2}{G_\alpha}^{\d\beta}.
\end{split}
\label{eq:Aabdot}
\end{equation}
In going to the second line, we have used 
$\cal{D}^{\rm P}_{\,\alpha}\Phi^i=\na_\alpha\Phi^i$ and 
$f_{\alpha\d\beta}=-G_{\alpha\d\beta}/2$ which reads from the
curvature constraints after the gauge fixing. The second order
derivative is modified by using
\begin{equation}
\b{\cal{D}}^{{\rm P}\d\beta}\cal{D}^{\rm P}_\alpha \Phi^i
+iA^{\d\beta}\cal{D}^{\rm P}_\alpha \Phi^i
=\b\na^{\d\beta}\na_\alpha\Phi^i
=-2i\na_\alpha{}^{\d\beta}\Phi^i
=-2i\cal{D}^{\rm P}{}_\alpha{}^{\d\beta}\Phi^i,
\end{equation}
which also follows from (\ref{eq:N2formula}) and the gauge-fixing
conditions. 
Eq.~(\ref{eq:Aabdot}) is regarded as determining ${G_\alpha}^{\d\beta}$ 
in terms of the auxiliary $A$-gauge field $A_a$.

\subsection{Correspondence to component approach}

We here show the correspondence of superconformal gauge fixing between
the superspace and component approaches. First, note that the
correspondences of the potentials and compensators are as follows:
\begin{equation}
\begin{array}{c|c}
\hline\hline
\text{component} & \text{superspace}
\\ \hline
\,S_i,\ \ \Sigma_{\text{c}},\ \ \Sigma_0 \,
& \Phi^i,\ \ \Phi^\co,\ \ \Phi^0
\\ \hline
\Tspan{\tilde{\phi}},\ \ \phi
& \, 3e^{-K/3},\ \ 3e^{-G/3}
\\ \hline
g,\ \ \cal{G} 
& W,\ \ -G
\\ \hline
\end{array}
\end{equation}
The symbols in component approach are explained in section~\ref{sec:comapp}. 

Let us see that the gauge-fixing conditions (\ref{eq:GF-G}) in
superspace are equivalent to the improved $D$, $A$, $S$, $K_a$ gauge 
conditions (\ref{eq:KUgaugeCond}) in component approach. As discussed
in section~\ref{sec:proj-act}, the component correspondence between
the compensator multiplet $\Sigma_0$ and the compensator superfield
$\Phi^0$ is
\begin{equation}
\Sigma_0=[\,z_0,\ \cal{P}_{\text{R}}\chi_0,\ h_0\,] 
\quad\corr\quad
\big[\, \Phi^0|,\ \ \na_\alpha\Phi^0|,\ \ -\tfrac14\na^2\Phi^0|\,\big].
\end{equation}
The gauge conditions (\ref{eq:GF-G}) or its consequence
(\ref{eq:phi0}) directly means the correspondence of the gauge-fixed 
lowest components
\begin{equation}
z_0 = \sqrt{3} \phi^{-\frac{1}{2}}(z,z^*) = e^{-\cal{G}/6} 
\quad\leftrightarrow\quad
\Phi^0\big| = e^{G/6}\big|.
\end{equation}
For the spinor components, the $S$-gauge condition in component
approach and Eq.~(\ref{eq:fix_spinor}) in superspace exactly agree with
each other:
\begin{equation}
2\chi_{\text{R}0} =
 -2z_0\phi^{-1}\phi^i\chi_{\text{R}i} =
-\df{1}{3}e^{-\cal{G}/6}\cal{G}^i (2 \chi_{\text{R}i})
\quad\leftrightarrow\quad
\na_\alpha\Phi^0\big| = \df{1}{3}e^{G/6}G_i \nabla_\alpha\Phi^i\big|.
\label{eq:spi_corr}
\end{equation}
The correspondence of the $K_a$ gauge is trivial. Note that the $S$
gauge condition in superspace approach, $B_\alpha=0$, is used in
deriving the spinor component of $\Phi^0$ (\ref{eq:fix_spinor}), which
leads to the correspondence (\ref{eq:spi_corr}). For the $F$
components, the auxiliary field $h_0$ in $\Sigma_0$ is not gauge-fixed
in component approach. This corresponds to the fact in superspace that 
the $F$ component of $\Phi^0$ contains the auxiliary part $\b{R}$
after the gauge fixing, as given in (\ref{eq:hcompensator}).

The virtue of gauge fixing in conformal superspace is two fold. The
first is that it leads to the superspace Poincar\'e SUGRA directly and
easily. The second is that it finds the supersymmetry transformation
in the resultant Poincar\'e SUGRA in a straightforward way. Namely,
the remaining Poincar\'e supersymmetry is just given by the covariant
spinor derivatives $\cal{D}^{\rm P}_{\,\alpha}$ and 
$\b{\cal{D}}^{\rm P}_{\,\d\alpha}$. In component approach, however,
the remaining supersymmetry is deformed from the original one by the
requirement that it keeps the $D$, $A$, $S$, $K_a$ gauge conditions
intact, and explicitly found in Ref.~\cite{bib:KUigc} by adding a
complicated combination of the $A$, $S$, $K_a$ gauge transformations
with non-trivial field-dependent parameters (\ref{eq:omPQ}). We
finally show this correspondence of the Poincar\'e supersymmetry after
the gauge fixing, in particular, the covariant spinor derivative
$\cal{D}^{\rm P}_{\,\ul{\alpha}}$ reproduces the deformed
supersymmetry in component approach.

The Poincar\'e spinor derivative (\ref{eq:PoincareDer}) is related
to the conformal one as $\cal{D}^{\rm P}_{\,\ul{\alpha}}= 
\na_{\ul{\alpha}}+ A_{\ul{\alpha}}A + f_{\ul{\alpha}}{}^AK_A$ after
the gauge fixing. The supersymmetry transformation in Poincar\'e
superspace defined by $\eta^{\ul{\alpha}}\cal{D}^{\rm P}_{\,\ul{\alpha}}$ 
is then given by the following linear combination of superconformal
$A$, $K_A$ transformations
\begin{equation}
\begin{split}
\eta^{\ul{\alpha}}\cal{D}^{\rm P}_{\,\ul{\alpha}} 
&= \eta^{\ul{\alpha}}\na_{\ul{\alpha}} + \xi(A)'(\eta)A 
+ \xi(K)'(\eta)^{\ul{\alpha}}S_{\ul{\alpha}} + \xi(K)'(\eta)^a K_a, \\  
& \xi(A)'(\eta)\ = \eta^\alpha A_{\alpha} + \b\eta_{\d\alpha}A^{\d\alpha}
=\frac{i}4G_j \eta^\alpha \cal{D}_\alpha^{\rm P}\Phi^j 
-\frac{i}4G_{j*}\b{\eta}_{\d{\alpha}} \b{\cal{D}}^{{\rm P}\,\d{\alpha}}
\b{\Phi}^{j*},  \\ 
& \xi(K)'(\eta)_\alpha = \eta^\beta f_{\beta\alpha}+
\b\eta_{\d\beta}f^{\d\beta}{}_\alpha = \eta_\alpha\b{R} 
+\df12G_\alpha{}^{\d\beta}\b\eta_{\d\beta}, \\  
& \xi(K)'(\eta)^{\d\alpha} = \eta^\beta f_\beta{}^{\d\alpha}+
\b\eta_{\d\beta}f^{\d\beta\d\alpha} = -\b\eta^{\d\alpha} R
-\df12\eta^\beta G_\beta{}^{\d\alpha}, \\  
& \xi(K)'(\eta)^a = \eta^\beta f_{\beta}{}^a+ \b\eta_{\d\beta}f^{\d\beta a}.
\end{split}
\label{eq:xis}
\end{equation}
Similarly, the Poincar\'e $Q$ transformation after the gauge fixing is
written as
\begin{equation}
\delta_G(\eta^{\ul{\alpha}} Q^{\text{P}}_{\ul{\alpha}})
=\delta_G(\eta^{\ul{\alpha}}Q_{\ul{\alpha}})
+\delta_G(\xi(A)'(\eta)A) +\delta_G(\xi(K)'(\eta)^BK_B)
\end{equation}
with the same parameters given in (\ref{eq:xis}). We show this
transformation is exactly the same as the $Q$ transformation
(\ref{eq:omPQ}) in component approach by examining the correspondence
between the transformation parameters (\ref{eq:xis}) and
(\ref{eq:PQparameters}). For the $A$ transformation, the parameter in
superspace is
\begin{equation}
\xi(A)'(\eta)\big| = \frac34 \(\frac{i}3 G_j (2\eta^\alpha) 
\big(\tfrac{1}{2}\na_\alpha\Phi^j\big)
-\frac{i}3 G_{j^*}(2\b{\eta}_{\d{\alpha}} )
\big(\tfrac{1}{2}\b\na^{\d{\alpha}}\b\Phi^{j^*}\big) \)\!\Big| 
\label{eq:xiAeta}
\end{equation}
Noticing the parameter correspondences 
$\tfrac{3}{4}\theta\leftrightarrow\xi(A)|$ 
and $\b\varepsilon \corr 2\bmx\eta^\alpha & \b\eta_{\d\alpha}\emx\!|$ 
given in Table~(\ref{eq:parameter}), we find (\ref{eq:xiAeta}) agrees
with $\theta(\varepsilon)$ of (\ref{eq:PQparameters}) in component
approach. For the $S$ transformation, the above parameter
$\xi(K)'(\eta)_\alpha$ in superspace is rewritten by the other
auxiliary fields with (\ref{eq:hcompensator}) and (\ref{eq:Aabdot}),
and given by
\begin{equation}
\begin{split}
& \xi(K)'(\eta)_\alpha \big| = \Big(\eta_\alpha \b{R} 
+\frac{1}{2}\b\eta_{\d\beta}G_\alpha{}^{\d\beta}\Big)\Big|  \\[1mm]
&\quad =\frac{-1}{2} \bigg\{-\df{1}{2} 
\Bigl(\(-\tfrac{1}{4}\nabla^2 \Phi^0\)e^{-G/6}
-\df{1}{3}\(-\tfrac{1}{4}\nabla^2\Phi^i\)G_i\Bigr)(2\eta_\alpha)  \\
&\hspace*{15mm}
+\df{1}{3}\( \bigl( G_{ij}+\df{1}{3}G_iG_j\bigr) 
(2\eta^\gamma) \( \tfrac{1}{2}\na_\gamma \Phi^j \)
+G_{ij^*} (2\b\eta_{\d\beta})
\big( \tfrac{1}{2}\b\na^{\d\beta}\b\Phi^{j^*} \big) \)
\( \tfrac{1}{2}\na_\alpha\Phi^i \)  \\
&\hspace*{15mm}
+\df{1}{12}{e_c}^m \( G_i\na_m \Phi^i -G_{i^*}\na_m\b\Phi^{i^*} \)
(i\sigma^c_{\alpha\d\beta}(2\b\eta^{\d\beta}))
+\df{i}{4}{e_c}^m \(\tfrac{4}{3}A_m\)
(i\sigma^c_{\alpha\d\beta}(2\b\eta^{\d\beta})) \bigg\}\Big| ,
\end{split}
\end{equation}
which is the same as $\zeta_{\rm R}(\varepsilon)$ of
(\ref{eq:PQparameters}) in component approach with the correspondences
of parameters and gauge fields given in Tables~(\ref{eq:parameter}) and
(\ref{eq:GF}),
especially 
$\b\zeta\leftrightarrow -2(\xi(K)^\alpha,\b\xi(K)_{\d\alpha})|$
and
$\frac{3}{4}A_\mu\leftrightarrow A_m|$.
A similar
argument holds for $\xi(K)'(\eta)^{\d\alpha}$. Finally, we discuss the
$K_a$ transformation part. Using $f_{\beta a}=-f_{a\beta}$ which comes
from the curvature constraints after the gauge fixing, we have
\begin{equation}
f_{\beta a}| = -{e_a}^mf_{m\beta}|
+\df{1}{2}{e_a}^m{\psi_m}^{\alpha}f_{\alpha\beta}|
+\df{1}{2}{e_a}^m\b\psi_{m\d\alpha}{f^{\d\alpha}}_\beta| \,.
\end{equation}
With this form at hand, the parameter $\xi(K)'(\eta)_a$ of
(\ref{eq:xis}) in superspace is rewritten as
\begin{equation}
\begin{split}
(\eta^\beta f_{\beta a}+ \b\eta_{\d\beta}f^{\d\beta}{}_a)|
&= \df{1}{4}\Big\{ e_a{}^m\Bigl((-2{f_m}^\beta)(2\eta_\beta)
+(-2f_{m\d\beta})(2\b\eta^{\d\beta})\Bigr)  \\
&\hphantom{=\delta_G\Bigg(\df{1}{4}} -e_a{}^m\psi_m{}^\alpha
\bigl(-2\bigl(\eta^\beta f_{\beta\alpha}
+\b\eta_{\d\beta}{f^{\d\beta}}_\alpha\big)\bigr)  \\
&\hphantom{=\delta_G\Bigg(\df{1}{4}} -e_a{}^m\b\psi_{m\d\alpha}
\bigl( -2\bigl(\eta^\beta{f_\beta}^{\d\alpha}
+\b\eta_{\d\beta}{f^{\d\beta\d\alpha}}\big)\bigr) \Big\}\Big| \,.
\end{split}
\end{equation}
The RHS is same as $\xi_a(\varepsilon)$ of (\ref{eq:PQparameters}) 
in component approach with the correspondences of the $S$ gauge field
given in Table~(\ref{eq:GF}) and the $S$ transformation parameter
discussed above.

\section{Summary}

In this paper, we have investigated the four-dimensional $\cal{N}=1$
conformal SUGRA in two different approaches. One is the superconformal
tensor calculus, developed in 1980's~\cite{bib:KU} which uses the
ordinary four-dimensional field theory. The other is the superspace
formalism, recently constructed in~\cite{bib:B1}, which uses the
conformal superspace and superfields.

Though there are apparent difference in supersymmetry transformation
and superconformal multiplets, we have shown that two approaches are
completely equivalent, and clarified the correspondences of
superconformal generators (\ref{eq:SG}), gauge fields (\ref{eq:GF}),
(\ref{eq:corrgaugefield}), curvatures (\ref{eq:CurvatureCorr}),
(\ref{eq:corrRab}) and their constraints (read from (\ref{eq:R-W})), 
superconformal transformations (\ref{eq:SGT}), multiplet fields
(\ref{eq:KUBcorr}), chiral projection (\ref{eq:proj-corr}), 
and invariant actions (\ref{eq:F-corr}), (\ref{eq:DD-corr}).

The action in superspace formalism has a huge number of gauge
invariance than the component approach. Therefore the correspondence
between two approaches should be clarified also for the gauge fixing
conditions. We make comprehensible how to obtain Poincar\'e SUGRA and
the remaining supersymmetry by fixing the superconformal gauge
symmetry in the general matter-coupled SUGRA system.

\bigskip

\subsection*{Acknowledgments}
\noindent
The authors thank Shuntaro Aoki, Tetsutaro Higaki, Tetsuji Kimura,
Michinobu Nishida, Yusuke Yamada and Naoki Yamamoto for helpful
discussions and comments. T.K.\ also thanks Ariyoshi Kunitomo for his
discussions and supports for this work. 
This work is partially
supported by 
the Research Grant of Keio Leading-edge Laboratory of Science \& Technology.
The discussions
during the Yukawa Institute for Theoretical Physics workshop
YITP-W-15-20 on ``Microstructures of black holes'' were useful to
complete this work.

\newpage
\appendix
\section{Notations}
\label{sec:notation}

In the component approach part in the text, we use the notation of
KU~\cite{bib:KU}, which is the same as Ref.~\cite{bib:vN} except for
two-component spinors and the dual of second rank anti-symmetric
tensor. In the superspace approach part, we use the notation of Wess
and Bagger \cite{bib:WB}.

\subsection{Notation in component approach}

We use Roman letters for flat Lorentz indices, Greek letters
$\mu,\nu,\ldots$ for curved vectors, and Greek letters
$\alpha,\beta,\ldots$ for two-component spinors. We also use the
Euclidian notation (the Pauli metric). The metric and the totally
anti-symmetric tensor are given by
\begin{equation}
\delta_{ab}=\textrm{diag}(1,1,1,1), \qquad \varepsilon^{1234}=1.
\end{equation}
The gamma matrices satisfy
\begin{equation}
\{\gamma_a,\gamma_b\}=2\delta_{ab},
\end{equation}
and $\gamma_5$ and $\sigma_{ab}$ are defined as
\begin{equation}
\gamma_5=\gamma_1\gamma_2\gamma_3\gamma_4=
\bmx 1 & 0 \\ 0 & -1 \emx,
\qquad 
\sigma_{ab}=\df14(\gamma_a\gamma_b-\gamma_b\gamma_a)=
\bmx (\sigma_{ab})_\alpha{}^\beta & 0 \\
0 & (\b{\sigma}_{ab})^{\d{\alpha}}{}_{\d{\beta}} \emx.
\end{equation}
The relation between four-component and two-component spinors is
\begin{equation}
Psi=
\bmx \psi_{\alpha} \\ \psi^{\d\alpha}\emx =
\bmx \psi_{\text{R}}\\ \psi_{\text{L}} \emx,
\qquad
\b\Psi=\Psi^T C =
\bmx \psi^\alpha & \psi_{\d\alpha}\emx = 
\bmx \b\psi_{\text{R}} & \b\psi_{\text{L}} \emx,
\end{equation}
\begin{equation}
C = 
\bmx -\epsilon^{\alpha\beta} & 0 \\
0 & \epsilon_{\d\alpha\d\beta} \emx,  
\qquad
C^{-1} = 
\bmx \epsilon_{\alpha\beta} & 0 \\
0 & -\epsilon^{\d\alpha\d\beta} \emx,
\end{equation}
where $\epsilon^{\alpha\beta}$ is the anti-symmetric tensor with
$\epsilon^{12}=\epsilon_{12}=1$. The raising and lowering rules of
spinor index are defined by
\begin{equation}
\psi^\alpha = \epsilon^{\alpha\beta}\psi_\beta, \qquad
\psi_\alpha = \psi^\beta\epsilon_{\beta\alpha}.
\end{equation}

The dual of anti-symmetric tensor $F_{ab}$, and its self-dual and
anti-self-dual parts are defined as
\begin{equation}
\tilde{F}_{ab} := \df{1}{2}\varepsilon_{abcd}F^{cd},
\qquad  F^\pm_{ab} := \df{1}{2}(F_{ab}\pm \tilde{F}_{ab}).
\end{equation}
Using the relation $\tilde\sigma_{ab}= -\gamma_5\sigma_{ab}$, we find 
\begin{equation}
\sigma_{ab}\psi_{\rm R}= \sigma_{ab}^- \psi_{\rm R}, \qquad
\sigma_{ab}\psi_{\rm L}= \sigma_{ab}^+ \psi_{\rm L}, 
\qquad 
\sigma^{ab} F^{\pm}_{ab}=\df{1\mp \gamma_5}{2}\sigma^{ab}F_{ab}.
\end{equation}

\subsection{Notation in superspace approach}
\label{sec:ssnotation}

We use the indices $a,b,...$ for flat Lorentz vectors,
$\alpha,\beta,...$ for flat Lorentz spinors, $m,n,...$ for curved
vectors, and $\mu,\nu,...$ for curved spinors. The indices 
$A,B,\ldots$ are the sets of flat vectors and spinors, and
$M,N,\ldots$ the sets of curved vectors and spinors. We also use the 
Minkowski metric. The metric and the totally anti-symmetric tensor are
given by
\begin{equation}
\eta_{ab}= \text{diag}(-1,1,1,1), \qquad 
\varepsilon^{0123}=-\varepsilon_{0123}=1.
\end{equation}
The standard contractions of two-component spinors are
\begin{equation}
\xi\psi=\xi^\alpha\psi_\alpha, \qquad
\b\xi\b\psi=\b\xi_{\d\alpha}\b\psi^{\d\alpha},
\end{equation}
and the raising and lowering rules of index are defined by
\begin{equation}
\psi^\alpha = \epsilon^{\alpha\beta}\psi_\beta, \qquad 
\psi_\alpha = \epsilon_{\alpha\beta}\psi^\beta, \qquad
\b\psi^{\d\alpha} = \epsilon^{\d\alpha\d\beta}\b\psi_{\d\beta}, \qquad
\b\psi_{\d\alpha} = \epsilon_{\d\alpha\d\beta}\b\psi^{\d\beta},
\end{equation}
where $\epsilon^{\alpha\beta}$ and $\epsilon^{\d\alpha\d\beta}$ are
the second-order anti-symmetric tensors 
with $\epsilon^{12}=\epsilon_{21}=1$. The hermitian conjugate of
spinor is given by $(\psi_\alpha)^\dagger=\b\psi_{\d\alpha}$, and
the hermitian conjugate rule for spinor product is
\begin{equation}
(\xi_\alpha\psi_\beta)^\dagger = \b\psi_{\d\beta}\,\b\xi_{\d\alpha}.
\end{equation}
The four-dimensional Pauli matrices $\sigma_a$ are defined as
\begin{equation}
(\sigma_0,\,\sigma_1,\,\sigma_2,\,\sigma_3)_{\alpha\d\beta}
= \Big( \bmx 1 & 0 \\ 0 & 1 \emx,\,
\bmx 0 & 1 \\ 1 & 0 \emx,\,
\bmx 0 & -i \\ i & 0 \emx,\,
\bmx 1 & 0 \\ 0 & -1 \emx \Big),
\end{equation}
and their hermitian conjugates are
\begin{equation}
(\b\sigma_a)^{\d\alpha\beta}=\varepsilon^{\d\alpha\d\gamma}
\varepsilon^{\beta\delta} (\sigma_a)_{\delta\d\gamma} =
(\sigma_a)^{\beta\d\alpha}.
\end{equation}
With these matrices, any flat Lorentz vector $V_a$ can be expressed as
a mixed spinor $V_{\alpha\d{\beta}}$ and vice versa:
\begin{equation}
V_{\alpha\d{\beta}}= (\sigma^a)_{\alpha\d{\beta}}V_a, \qquad 
V^a= -\df12 \(\b{\sigma}^a\)^{\d{\beta}\alpha}
V_{\alpha\d{\beta}}\,.
\end{equation}
The matrices $\sigma^{ab}$ and $\b\sigma^{ab}$ are defined as
\begin{equation}
(\sigma^{ab})_\alpha{}^\beta=
\df{1}{4}(\sigma^a\b\sigma^b-\sigma^b\b\sigma^a)_\alpha{}^\beta, \qquad
(\b\sigma^{ab})^{\d\alpha}{}_{\d\beta}=
\df{1}{4}(\b\sigma^a\sigma^b-\b\sigma^b\sigma^a)^{\d\alpha}{}_{\d\beta},
\end{equation}
and satisfy the relations
\begin{equation}
\varepsilon_{abcd} {(\sigma^{cd})_\alpha}^\beta 
= -2i{(\sigma_{ab})_\alpha}^\beta, \qquad
\varepsilon_{abcd} {(\bar{\sigma}^{cd})^{\d\alpha}}_{\d\beta}
= 2i{(\b{\sigma}_{ab})^{\d\alpha}}_{\d\beta}\,.
\label{eq:sig_eps}
\end{equation}
In two-component spinor notation, any anti-symmetric tensor $F_{ab}$ can 
be decomposed into chiral and anti-chiral parts:
\begin{equation}
F_{ab} = -(\epsilon\sigma_{ab})^{\alpha\beta} F^-_{\alpha\beta}
+(\b{\sigma}_{ab}\epsilon)^{\d\alpha\d\beta} F^+_{\d\alpha\d\beta},  
\end{equation}
where 
\begin{equation}
F^-_{\alpha\beta} = \df{1}{2}(\sigma^{ab}\epsilon)_{\alpha\beta}F_{ab},
\qquad
F^+_{\d\alpha\d\beta} = -\df{1}{2}
(\epsilon\b\sigma^{ab})_{\d\alpha\d\beta}F_{ab}.
\label{eq:chiral_decom}
\end{equation}
The dual of anti-symmetric tensor $F_{ab}$ is defined as
\begin{equation}
(*F)_{ab} := \df{1}{2}\varepsilon_{abcd}F^{cd} =
i(\epsilon\sigma_{ab})^{\alpha\beta} F^-_{{\alpha\beta}} +
i(\b{\sigma}_{ab}\epsilon)^{\d\alpha\d\beta} F^+_{\d\alpha\d\beta}.
\end{equation}
The self-dual and anti-self-dual parts of $F_{ab}$ are
\begin{equation}
F^{\pm}_{ab}:=\df{1}{2}\bigl(F_{ab}\mp i(*F)_{ab}\bigr), 
\end{equation}
which coincide with the chiral and anti-chiral parts, respectively:
\begin{equation}
F^-_{ab}=-(\epsilon\sigma_{ab})^{\alpha\beta}F^-_{{\alpha\beta}},
\qquad
F^+_{ab}=(\b\sigma_{ab}\epsilon)^{\d\alpha\d\beta} F^+_{\d\alpha\d\beta}.
\end{equation}

\subsection{Correspondence of notations}

We summarize the correspondence of notations between component and
superspace approaches:
\begin{equation}
\begin{array}{c|c}
\hline\hline
\text{component} & \text{superspace} 
\\ \hline
\delta_{ab} & \eta_{ab}
\\ \hline
\Tspan{\epsilon^{\alpha\beta},\;\; \epsilon_{\alpha\beta}}  & 
\epsilon^{\alpha\beta},\;\; -\epsilon_{\alpha\beta} 
\\ \hline
\Tspan{\epsilon^{\d\alpha\d\beta},\;\; \epsilon_{\d\alpha\d\beta}} &
\epsilon^{\d\alpha\d\beta},\;\; -\epsilon_{\d\alpha\d\beta} 
\\ \hline
\Tspan{\bmx \psi_\alpha \\ \b\psi^{\d\alpha} \emx},\;\;
\Tspan{\bmx \psi^\alpha & \b\psi_{\d\alpha} \emx}  &  
\bmx \psi_\alpha \\ \b\psi^{\d\alpha} \emx, \;\;
\bmx \psi^\alpha & \b\psi_{\d\alpha} \emx
\\ \hline
\gamma_a & \;\Tspan{i\gamma_a=i\!\stog{a}{}{\alpha}{\beta}}
\\ \hline
\gamma_5 &
\bmx 1 & 0 \\ 0 & -1 \emx
\\ \hline
\sigma_{ab} & 
\Tspan{-\bmx
(\sigma_{ab})_\alpha{}^\beta & 0 \\
0 & (\bar{\sigma}_{ab})^{\d\alpha}{}_{\d\beta}
\emx}
\\ \hline
\varepsilon_{abcd} & -i\varepsilon_{abcd}
\\ \hline
\Tspan{\tilde{F}_{ab}} 
& -i(*F)_{ab} 
\\ \hline
F^{\pm}_{ab} 
& F^{\pm}_{ab} 
\\ \hline
\end{array}
\label{eq:notation}
\end{equation}

\section{$Q$ transformation of conformal multiplet}

The supersymmetry $Q$ transformation laws take the following form 
for the fields in a general conformal 
multiplet $[\cal{C}_\Gamma,\cal{Z}_\Gamma,\cal{H}_\Gamma,
\cal{K}_\Gamma,\cal{B}_{a\Gamma},\Lambda_\Gamma,\cal{D}_\Gamma]$:
\begin{equation}
\begin{split}
\delta_Q(\varepsilon)\cal{C}_\Gamma & =
\df{1}{2}i\b\varepsilon \gamma_5 \cal{Z}_\Gamma,  \\
\delta_Q(\varepsilon) \cal{Z}_\Gamma & =
(-)^\Gamma \df{1}{2} (i\gamma_5 \cal{H}_\Gamma-\cal{K}_\Gamma 
-\gamma^a\cal{B}_{a\Gamma}
+i\gamma^a D_a\cal{C}_\Gamma \gamma_5) \varepsilon,  \\
\delta_Q(\varepsilon)\cal{H}_\Gamma & = 
\df{1}{2}i\b{\varepsilon}\gamma_5
(\gamma^a D_a\cal{Z}_\Gamma+\Lambda_\Gamma),\\
\delta_Q(\varepsilon)\cal{K}_\Gamma & =
-\df{1}{2}\b{\varepsilon}(\gamma^a D_a\cal{Z}_\Gamma+\Lambda_\Gamma), \\
\delta_Q(\varepsilon)\cal{B}_{a\Gamma} & =
-\df{1}{2}\b{\varepsilon}(D_a\cal{Z}_\Gamma+\gamma_a\Lambda_\Gamma)
-\df{1}{4}iR_{bc}(Q)\gamma_5\gamma_a(\Sigma_{bc}\cal{C})_\Gamma,  \\
\delta_Q(\varepsilon) \Lambda_\Gamma & = 
(-)^\Gamma\df{1}{2} (\sigma^{ab}\cal{F}_{ab\,\Gamma}+
i\gamma_5 \cal{D}_\Gamma)\varepsilon  \\
& \qquad +\df{1}{8}\(
\gamma_c \varepsilon R_{ab}(Q)\gamma_c (\Sigma^{ab} \cal{Z})_\Gamma 
+\gamma_5 \gamma_c\varepsilon R_{ab} (Q) \gamma_5\gamma_c 
(\Sigma^{ab} \cal{Z})_\Gamma \),  \\
\delta_Q(\varepsilon)\cal{D}_\Gamma & =
\df{1}{2}i\b{\varepsilon}\gamma_5 \gamma^a D_a\Lambda_\Gamma 
-\df{1}{4}\b{\varepsilon}(R_{ab}(A)+\gamma_5\tilde{R}_{ab}(A))
(\Sigma^{ab} \cal{Z})_\Gamma  \\
&\qquad +(-)^\Gamma \df{1}{4}\bar{\varepsilon}
\big(i\gamma_5(\Sigma^{ab}\gamma^c\cal{B}_c)_\Gamma
-(\Sigma^{ab}\gamma^c D_c\cal{C})_\Gamma\big)
(R_{ab}(Q)C^{-1})^T.
\end{split} 
\label{eq:sf}
\end{equation}
This transformation law is called the standard form. The definition of
$\cal{F}_{ab\,\Gamma}$ in the transformation law of $\Lambda_\Gamma$ is
\begin{equation}
\cal{F}_{ab\,\Gamma} = D_a\cal{B}_{b\Gamma}-D_b\cal{B}_{a\Gamma}
+\df{1}{2}i\varepsilon_{abcd}[D^c,D^d]\cal{C}_\Gamma.
\label{eq:Fab}
\end{equation}

\section{Derivations of correspondence}
\label{sec:deriv}

\subsection{Conformal multiplets with arbitrary Lorentz indices}
\label{sec:deriv_multi}

In this subsection we explicitly derive the correspondences of 
conformal multiplets with arbitrary Lorentz index, that is, between 
$\cal{V}_\Gamma$ in component approach (Eq.~(\ref{eq:cfmatter})) and
$\Phi_\Gamma$ in superspace approach (Eq.~(\ref{eq:Bpsf})). In the 
first place, the correspondence of the lowest components is obtained
by the property of superconformal transformations. There is an
ambiguity for overall constant factor, which is fixed by
\begin{equation}
\cal{C}_\Gamma \ \ \corr\ \ \Phi_\Gamma|.
\end{equation}
We then obtain the correspondences of higher components by operating
the $Q$ transformations in order. The action of $Q$ transformation is
given by the covariant spinor derivative since the fields have only
Lorentz indices. As given in Table (\ref{eq:parameter}), the
correspondence of $Q$ transformation parameters is 
$\b\varepsilon\,\corr\, 2\bmx \xi(P)^\alpha & \b\xi(P)_{\d\alpha}\emx|$.
In the following, we simply denote the parameters in superspace as
$\bmx \xi^\alpha & \b\xi_{\d\alpha} \emx$.

The correspondence of the second components is obtained by the $Q$
transformations of the first components, namely,
\begin{equation}
\delta_Q(\varepsilon)\cal{C}_\Gamma \ \ \corr\ \
 (\xi^\alpha \nabla_\alpha + 
\b\xi_{\d\alpha}\b\nabla^{\d\alpha})\Phi_\Gamma 
= \df{1}{2}i
\bmx
2\xi^\alpha & 2\b\xi_{\d\alpha}
\emx
\axg
\bmx -i\na_\alpha\Phi_\Gamma \\ +i\b\na^{\d\alpha}\Phi_\Gamma\emx.
\end{equation}
By comparing with $\delta_Q(\varepsilon)\cal{C}_\Gamma$ in
(\ref{eq:sf}) and using the correspondence of $\gamma_5$ in 
Table (\ref{eq:notation}), we find
\begin{equation}
\cal{Z}_\Gamma \ \ \corr\ \
\bmx -i\na_\alpha\Phi_\Gamma \\ 
+i\b\na^{\d\alpha}\Phi_\Gamma\emx\!\Big|.
\label{eq:Z-corr}
\end{equation}
The correspondences of the other components are obtained in similar
ways. The $Q$ transformations of the second components are
\begin{equation}
\begin{split}
\delta_Q(\varepsilon)\cal{Z}_\Gamma \ \ \corr \ \ \ &
(\xi^\alpha \na_\alpha + \b\xi_{\d\alpha}\b\na^{\d\alpha})
\bmx -i\na_\beta\Phi_\Gamma \\ +i\b\na^{\d\beta}\Phi_\Gamma\emx  \\
& =(-)^{\Gamma}\df{1}{2}\(\,
\df{1}{4}(\na^2\Phi_\Gamma+\b\na^2\Phi_\Gamma )\,i
\bmx
\delta_\beta{}^\alpha & 0 \\
0 & -\delta^{\d\beta}{}_{\d\alpha} 
\emx  \right.  \\
& \qquad\qquad\quad
-\df{1}{4}(-i) (\na^2\Phi_\Gamma-\b\na^2\Phi_\Gamma )
\bmx
\delta_{\beta}{}^{\alpha} & 0 \\
0 & \delta^{\d\beta}{}_{\d\alpha} 
\emx  \\
& \qquad\qquad\quad
-\Big(-\df{1}{4}(\b{\sigma}_c)^{\gamma\d\gamma}
[\na_\gamma,\b\na_{\d\gamma}]\Phi_\Gamma\Big) \,i
\stog{}{c}{\beta}{\alpha}  \\
& \qquad\qquad \quad\left. 
+i\stog{}{c}{\beta}{\gamma} \na_c \Phi_\Gamma \,i
\bmx
\delta_\gamma{}^\alpha & 0 \\
0 & -\delta^{\d\gamma}{}_{\d\alpha} 
\emx  \, \)
\bmx 2\xi_\alpha \\ 2\b\xi^{\d\alpha} \emx.
\end{split}
\end{equation}
By comparing with $\delta_Q(\varepsilon)\cal{Z}_\Gamma$ in
(\ref{eq:sf}) and using the correspondence of gamma matrices in 
Table (\ref{eq:notation}), we find the correspondences of 
$\cal{H}_\Gamma$, $\cal{K}_\Gamma$, $\cal{B}_{a\Gamma}$ as
\begin{equation}
\begin{split}
& \cal{H}_\Gamma \ \ \corr\ \ 
\df{1}{4}(\na^2 \Phi_\Gamma + \b\na^2 \Phi_\Gamma)|,
\qquad
\cal{K}_\Gamma \ \ \corr\ \ 
-\df{1}{4}i(\na^2 \Phi_\Gamma - \b\na^2 \Phi_\Gamma)|,  \\
& \cal{B}_{a\Gamma} \ \ \corr\ \
-\df{1}{4}(\b\sigma_a)^{\d\beta\beta}
[\na_\beta,\b\na_{\d\beta}]\Phi_\Gamma|.
\end{split}
\label{eq:HKB}
\end{equation}

The $Q$ transformation of $\cal{H}_\Gamma$ implies
\begin{equation}
\begin{split}
\delta_Q(\varepsilon)\cal{H}_\Gamma \ \ \corr \ \ \ &
(\xi^\alpha \na_\alpha + \b\xi_{\d\alpha}{\b\na}^{\d\alpha})
\(\df{1}{4}(\na^2\Phi_\Gamma+\b\na^2\Phi_\Gamma)\)  \\
&=\df{1}{2}i
\bmx
2\xi^\alpha &2\b\xi_{\d\alpha}
\emx
\axg \( i\stog{}{c}{\alpha}{\beta}\na_c 
\bmx
-i\na_\beta\Phi_\Gamma \\
+i\b\na^{\d\beta}\Phi_\Gamma
\emx  \right.  \\
& \hspace*{55mm} \left.
+\df{1}{4}i
\bmx
-\b\na^2\na_\alpha \Phi_\Gamma+8\cal{W}_\alpha \Phi_\Gamma \\
+\na^2\b\na^{\d\alpha}\Phi_\Gamma+8\cal{W}^{\d\alpha}\Phi_\Gamma
\emx  \,\) .
\end{split}
\end{equation}
Here we have used the identities
\begin{equation}
\begin{split}
\na_\alpha\b\na^2 -\b\na^2 \na_\alpha  
+4i\na_{\alpha\d\beta}\bar\na^{\d\beta} +8\cal{W}_\alpha &=0,  \\
\b\na^{\d\alpha}\na^2 -\na^2\b\na^{\d\alpha} 
+4i\na^{\d\alpha\beta}\na_\beta -8\cal{W}^{\d\alpha} &=0 .
\end{split}
\label{eq:tnid}
\end{equation}
These identities are shown by evaluating (anti-)commutators of
covariant derivatives. By comparing with 
$\delta_Q(\varepsilon)\cal{H}_\Gamma$ in (\ref{eq:sf}), we find 
\begin{equation}
\Lambda_\Gamma \ \ \corr\ \ \df{i}{4}
\bmx
-\b\na^2\na_\alpha \Phi_\Gamma \\
+\na^2\b\na^{\d\alpha}\Phi_\Gamma
\emx\!\Big|
+2i
\bmx 
\cal{W}_\alpha \\ 
\cal{W}^{\d\alpha} 
\emx \Phi_\Gamma|,
\label{eq:lambda-corr}
\end{equation}
The correspondence of the $\Lambda_\Gamma$ component can also be
obtained by the $Q$ transformation of $\cal{K}_\Gamma$ or
$\cal{B}_{a\Gamma}$, which is found to be consistent with
(\ref{eq:lambda-corr}).

Finally, the $Q$ transformation of $\Lambda_\Gamma$ leads to the
correspondence of the $\cal{D}_\Gamma$ component. For the undotted
spinor in $\Lambda_\Gamma$, the $Q$ transformation is found from
(\ref{eq:lambda-corr}),
\begin{equation}
\begin{split}
\delta_Q(\varepsilon){\Lambda_\Gamma}_\alpha \ \ \corr \ \ \ &
(\xi^\beta \na_\beta+\b\xi_{\d\beta}\b\na^{\d\beta})
\(\df{-i}{4}\b\na^2\na_\alpha \Phi_\Gamma+2i\cal{W}_\alpha \Phi_\Gamma\)  \\
& =\xi^\beta (\sigma^{ab})_{\beta\alpha}
\( \na_a B_{b\Gamma}-\na_bB_{a\Gamma}-i[\na_a,\na_b]\Phi_\Gamma \)  \\
& \qquad\qquad
+i\xi_\alpha \(\df{1}{8}\na^\beta\b\na^2\na_\beta\Phi_\Gamma
-\cal{W}^\beta\na_\beta\Phi_\Gamma\)  \\
& \qquad\qquad
+2i\b\xi_{\d\beta}(R(P)_{cd})_\alpha M^{dc}\b\na^{\d\beta}\Phi_\Gamma.
\end{split}
\end{equation}
In the second line, we have introduced the superfield $B_{a\Gamma}$ 
made from the original $\Phi_\Gamma$ as
\begin{equation}
B_{a\Gamma} = -\df{1}{4}(\b\sigma_a)^{\d\beta\beta}
[\na_\beta,\b\na_{\d\beta}]\Phi_\Gamma,
\end{equation}
whose lowest component matches $\cal{B}_{a\Gamma}$ in component
approach as given in (\ref{eq:HKB}). For the dotted spinor 
in $\Lambda_\Gamma$, a similar expression holds. We then have the
correspondence
\begin{equation}
\begin{split}
\delta_Q(\varepsilon)\Lambda_\Gamma \ \corr \ \ &
(\xi^\beta\na_\beta+\b\xi_{\d\beta}\b\na^{\d\beta})
\(\df{i}{4}
\bmx
-\b\na^2\na_\alpha \Phi_\Gamma \\
+\na^2\b\na^{\d\alpha}\Phi_\Gamma
\emx
+2i \bmx
\cal{W}_\alpha \\
\cal{W}^{\d\alpha}
\emx
\Phi_\Gamma \)  \\
& = (-)^\Gamma\df{1}{2} (-1)
\bmx
{(\sigma^{ab})_\alpha}^\gamma & 0 \\
0 & {(\b\sigma^{ab})^{\d\alpha}}_{\d\gamma}
\emx
\(\na_a B_{b\Gamma}-\na_b B_{a\Gamma} 
+\df{1}{2}\varepsilon_{abcd} [\nabla^c,\na^d]\Phi_\Gamma\)
\bmx
2\xi_\gamma \\
2\b\xi^{\d\gamma}
\emx  \\
& \qquad
+(-)^\Gamma \df{1}{2}i\axg 
\(\df{1}{8}\b\na_{\d\beta}\na^2 \b\na^{\d\beta}\Phi_\Gamma
+\cal{W}_{\d\beta}\b\na^{\d\beta}\Phi_\Gamma  \)
\bmx
2\xi_\alpha \\
2\b\xi^{\d\alpha}
\emx  \\
& \qquad
+\df{1}{8}i\stog{c}{}{\alpha}{\beta}
\bmx
2\xi_\beta \\
2\b\xi^{\d\beta}
\emx  \\
& \qquad\qquad
\times (-2)
\bmx
(R(P)_{ab})^\delta & (R(P)_{ab})_{\d\delta}
\emx
i\stog{}{c}{\delta}{\gamma}(-M^{ab})
\bmx
-i\na_\gamma \Phi_\Gamma \\
+i\b\na^{\d\gamma}\Phi_\Gamma
\emx  \\
& \qquad
+\df{1}{8} i\stogm{c}{}{\alpha}{\beta}
\bmx
2\xi_\beta \\
2\b\xi^{\d\beta}
\emx  \\
& \qquad\qquad
\times (-2)
\bmx
(R(P)_{ab})^\delta & (R(P)_{ab})_{\d\delta}
\emx
i\stogm{}{c}{\delta}{\gamma}(-M^{ab})
\bmx
-i\na_\gamma \Phi_\Gamma \\
+i\b\na^{\d\gamma}\Phi_\Gamma
\emx.
\end{split}
\end{equation}
In this modification, we have used the relations (\ref{eq:sig_eps}),
the identity (\ref{eq:PTid}), and the equation
$\{\b\na_{\d\alpha},\cal{W}^{\d\alpha}\}\Phi_\Gamma=
-\tfrac12\na^\alpha \na^{\beta\d\gamma}
{W_{\beta\alpha}}^\gamma K_{\gamma\d\gamma}\Phi_\Gamma=0$ for a
primary superfield $\Phi_\Gamma$. By comparing with 
$\delta_Q(\varepsilon)\Lambda_\Gamma$ in (\ref{eq:sf}) and the 
definition (\ref{eq:Fab}), we find 
\begin{equation}
\cal{D}_\Gamma \ \ \corr\ \ 
\df{1}{8}\b\na_{\d\alpha}\na^2 \b\na^{\d\alpha}\Phi_\Gamma |
+\cal{W}_{\d\alpha}\b\na^{\d\alpha} \Phi_\Gamma |  \\
\,= \df{1}{8} \na^\alpha \b\na^2 \na_\alpha \Phi_\Gamma |
- \cal{W}^\alpha \na_\alpha \Phi_\Gamma |,
\label{eq:D-corr}
\end{equation}
and the correspondence 
\begin{equation}
\cal{F}_{ab\,\Gamma} \ \ \corr\ \ 
\(\na_a B_{b\Gamma}-\na_b B_{a\Gamma} \)|
+\df{1}{2}\varepsilon_{abcd} [\na^c,\na^d]\Phi_\Gamma|.
\end{equation}
That completes the correspondence of the components in a conformal
multiplet, which is summarized in Table~(\ref{eq:KUBcorr}).

The $Q$ transformation of $\cal{D}_\Gamma$ is a non-trivial
consistency check and explicitly calculated as
\begin{equation}
\begin{split}
\delta_Q(\varepsilon)\cal{D}_\Gamma \ \corr \ \ &
(\xi^\alpha\na_\alpha+\b\xi_{\d\alpha}\b\na^{\d\alpha})
\( \df{1}{8}\b\na_{\d\beta}\na^2 \b\na^{\d\beta}\Phi_\Gamma
+\cal{W}_{\d\beta}\b\na^{\d\beta} \Phi_\Gamma \)  \\
& = \df{1}{2}i
\bmx
2\xi^\alpha & 2\b\xi_{\d\alpha}
\emx
\axg i\stog{}{a}{\alpha}{\beta}\na_a
\( \df{i}{4}
\bmx
-\b\na^2\na_\beta\Phi_\Gamma \\
+\na^2\b\na^{\d\beta}\Phi_\Gamma
\emx +2i
\bmx
\cal{W}_\beta \\
\cal{W}^{\d\beta}
\emx\Phi_\Gamma
\)  \\
& \qquad
-\df{1}{4}
\bmx
2\xi^\alpha & 2\b\xi_{\d\alpha}
\emx
\(-\df{4}{3}\) \( R(A)_{ab}+\axg(-i)(*R(A))_{ab} \) (-M^{ab})
\bmx
-i\na_\alpha\Phi_\Gamma \\
+i\b\nabla^{\d\alpha}\Phi_\Gamma
\emx  \\
& \qquad
+\df{1}{4}(-)^\Gamma
\bmx
2\xi^\alpha & 2\b\xi_{\d\alpha}
\emx
i\axg \biggl(-M^{ab} \,i\stog{}{c}{\alpha}{\beta}B_{c\Gamma}\biggr)(-2)
\bmx
(R(P)_{ab})_\beta \\
(R(P)_{ab})^{\d\beta}
\emx  \\
& \qquad
-\df{1}{4}(-)^\Gamma
\bmx
2\xi^\alpha & 2\b\xi_{\d\alpha}
\emx \biggl(-M^{ab} \,i\stog{}{c}{\alpha}{\beta}\na_c\Phi_\Gamma\biggr)(-2)
\bmx
(R(P)_{ab})_\beta \\
(R(P)_{ab})^{\d\beta}
\emx.
\end{split}
\end{equation}
In this modification, we have used the relations
$\{\b\na_{\d\beta},\cal{W}^{\d\beta}\}\na_\alpha \Phi_\Gamma=
\tfrac{-1}{2}\na^\beta \na^{\gamma\d\gamma}
{W_{\gamma\beta}}^\delta K_{\delta\d\gamma}\na_\alpha\Phi_\Gamma=0$ and 
$\cal{W}(K)_{\d\beta}{}^cK_c\bar\na^{\d\beta}\na_\alpha\Phi_\Gamma=0$. 
Noticing the correspondences of generators, curvatures, gamma matrices,
lower components given in section~\ref{sec:corr} and above, we find that 
this form exactly agrees with $\delta_Q(\varepsilon)\cal{D}_\Gamma$
in (\ref{eq:sf}).

\subsection{Chiral projection}
\label{sec:deriv_proj}

In this subsection, we show the correspondence of the chiral projection
between two approaches. In component approach, the chiral projection
operator $\Pi$ acts on a  conformal multiplet $\cal{V}_\Gamma$ with
special weights and index, and gives a chiral multiplet
$\Pi\cal{V}_\Gamma$ whose component expression is explicitly given in
(\ref{eq:chiralproj}). In superspace approach, the chiral projection
operator $\cal{P}$ is defined by the superconformal covariant
derivative as $\cal{P}=\frac{-1}{4}\b\na^2$ and gives a chiral
superfield $\cal{P}\Phi_\Gamma$ from a primary superfield
$\Phi_\Gamma$ with special weights and index. In particular, $\Gamma$ 
should be made of purely undotted spinor indices. When one matches
$\cal{V}_\Gamma$ with $\Phi_\Gamma$, the correspondence of the chiral
projection is
\begin{equation}
\Pi\cal{V}_\Gamma \ \ \corr\ \ -\cal{P}\Phi_\Gamma =
\frac{1}{4}\b\na^2\Phi_\Gamma.
\end{equation}
In what follows, we show this correspondence explicitly by component
level, namely, each component of the chiral superfield 
$\frac{1}{4}\b\na^2\Phi_\Gamma$ coincide with (\ref{eq:chiralproj}) in
component approach.

For a general chiral superfield, its components which should match to
those of the corresponding chiral multiplet are given in
(\ref{eq:chiral-corr}). First, the lowest component of 
$\frac{1}{4}\b\na^2\Phi_\Gamma$ is 
\begin{equation}
\df{1}{4}\b\na^2\Phi_\Gamma|=
\df{1}{2}\(\df{1}{4}(\na^2\Phi_\Gamma+\b\na^2\Phi_\Gamma)|
-i\Big(-\df{1}{4}i(\na^2\Phi_\Gamma-\b\na^2\Phi_\Gamma)|\Big)\).
\end{equation}
We find from (\ref{eq:HKB}) that the RHS just corresponds to
$\frac12 \(\cal{H}_\Gamma-i\cal{K}_\Gamma\)$ in component approach,
which is the lowest component of the chiral multiplet
$\Pi\cal{V}_\Gamma$ as shown in (\ref{eq:chiralproj}).

The second component of chiral superfield is given by its covariant
derivative, and for $\frac{1}{4}\b\na^2\Phi_\Gamma$, it becomes
\begin{equation}
\na_\alpha\Big(\df14 \b\na^2 \Phi_\Gamma\Big)\Big| = 
i\Big( i(\sigma^a)_{\alpha\d\beta}\na_a(i\b\na^{\d\beta}\Phi_\Gamma)|
-\df{i}{4}\b\na^2\na_\alpha\Phi_\Gamma| +2i\cal{W}_\alpha\Phi_\Gamma| \Big),
\end{equation}
where the identity (\ref{eq:tnid}) has been used. By comparing with
the component correspondences (\ref{eq:Z-corr}) and
(\ref{eq:lambda-corr}), we find the RHS reads 
$i\cal{P}_\text{R}(\gamma^a D_a\cal{Z}_\Gamma+\Lambda_\Gamma)$ in
component approach, which is exactly the second component of
$\Pi\cal{V}_\Gamma$ given in (\ref{eq:chiralproj}). 

Finally the highest component of the chiral superfield
$\frac{1}{4}\b\na^2\Phi_\Gamma$ is
\begin{equation}
-\df{1}{4}\na^2\Big(\df14\b\na^2\Phi_\Gamma\Big)\Big| = 
-\df{1}{2}\( \df{1}{8}\b\na_{\d\alpha}\na^2\b\na^{\d\alpha}\Phi_\Gamma|
+\na^a\na_a \Phi_\Gamma|
+i\na_a \Bigl(-\df{1}{4}(\b\sigma^a)^{\d\alpha\alpha}
[\na_\alpha,\b\na_{\d\alpha}]\Phi_\Gamma\Bigr)\big| \),
\label{eq:Fcorr}
\end{equation}
where we have used the identity (\ref{eq:ChiralIdentity}), and 
the equation $\cal{W}_{\d{\alpha}}\b{\na}^{\d{\alpha}}\Phi_\Gamma=0$
which holds on a primary $\Phi_\Gamma$ with purely undotted $\Gamma$,
namely, $\cal{W}_{\d{\alpha}}\b{\na}^{\d{\alpha}}\Phi_\Gamma=
\{\b\na^{\d\alpha},\cal{W}_{\d\alpha}\}\Phi_\Gamma=
\frac{1}{2}\na^\alpha\na^{\beta\d\gamma}{W_{\beta\alpha}}^\gamma
K_{\gamma\d\gamma}\Phi_\Gamma=0$ since 
$\cal{W}_{\d\alpha}\Phi_\Gamma=0$ for such $\Phi_\Gamma$. By comparing
with the correspondence of $\cal{B}_{a\Gamma}$ in (\ref{eq:HKB}) and
also (\ref{eq:D-corr}) noting again 
$\cal{W}_{\d{\alpha}}\b{\na}^{\d{\alpha}}\Phi_\Gamma=0$, 
the RHS of (\ref{eq:Fcorr}) corresponds to 
$-\tfrac12(\cal{D}_\Gamma+\square\cal{C}_\Gamma+iD^a\cal{B}_{a\Gamma})$
in component approach, which is the highest $\cal{F}$ component of
$\Pi\cal{V}_\Gamma$ as shown in (\ref{eq:chiralproj}).

\end{document}